\documentclass[authoryear,12pt,onecolumn,3p]{elsarticle}
\usepackage{amssymb}
\usepackage{amsmath}
\usepackage{amsfonts}
\usepackage{amsbsy}
\usepackage{graphicx}
\usepackage{pifont}
\usepackage[authoryear]{natbib}
\usepackage{geometry}
\usepackage{hyperref}
\usepackage{version}
\usepackage{setspace}
\providecommand{\U}[1]{\protect\rule{.1in}{.1in}}
\journal{to arXiv}
\newtheorem{theorem}{Theorem}
\newtheorem{proposition}[theorem]{Proposition}

\newtheorem{remark}[theorem]{Remark}

\newtheorem{lemma}[theorem]{Lemma}

\newtheorem{example}[theorem]{Example}
\excludeversion{comment2}
\excludeversion{comment3}
\begin{document}

\begin{frontmatter}
\title{Balanced contributions, consistency, and value for games with externalities}
\author[hhl]{Andr{\'{e}} Casajus}
\ead{mail@casajus.de}
\ead[url]{www.casajus.de}
\author[was]{Yukihiko Funaki}
\ead{funaki@waseda.jp}
\ead[url]{yfunaki.blogspot.com}
\author[skk]{Frank Huettner\corref{cor}}
\ead{mail@frankhuettner.de}
\ead[url]{www.frankhuettner.de}
\cortext[cor]{Corresponding author}
\address[hhl]{\sl HHL Leipzig Graduate School of Management, Jahnallee~59, 04109~Leipzig, Germany}
\address[was]{\sl School of Political Science and Economics, Waseda University, Nishi-Waseda, Shinjuku-Ku, Tokyo, Japan}
\address[skk]{\sl SKK GSB, Sungkyunkwan University, 25-2, Sungkyunkwan-Ro, Jongno-gu, 110-745 Seoul, South Korea}
\begin{abstract}
We consider fair and consistent extensions of the Shapley value for games with externalities. Based on the restriction identified by Casajus et al. (2024, Games Econ. Behavior 147, 88-146), we define balanced contributions, Sobolev's consistency, and Hart and Mas-Colell's consistency for games with externalities, and we show that these properties lead to characterizations of the generalization of the Shapley value introduced by Macho-Stadler et al. (2007, J. Econ. Theory 135, 339-356), that parallel important characterizations of the Shapley value.
\end{abstract}
\end{frontmatter}

\section{Introduction}

The question of how to divide jointly created value in cooperative games has
been fundamentally influenced by the concept of the Shapley value. This
solution concept was originally designed to capture a player's value in a
cooperative game with transferable utility (henceforth TU game), but its
application nowadays transitions into seemingly unrelated domains such as
statistics for identifying important variables
{\citep{LipCon2001,Shorrocks2012}}\ or into machine learning for interpreting
prediction models {\citep{LunLee2017,Lundbergetal2020}}. The Shapley value
applies under the presupposition that the worth of a coalition is independent
of the coalition structure of the outside players. However, this assumption is
challenged when external effects come into play, that is, when the actions of
one coalition exert influence over another's outcomes. To capture
externalities,
\citet{ThrLuc1963}%
\ introduced the framework of \textit{games with externalities} (henceforth
\textit{TUX games}). Crucially, in a TUX game, the worth of a coalition may
depend on the coalitions formed by the outsiders, which makes it a
generalization of TU games.

TUX games are often used---but are not limited---to study coalition formation
\citep{GraFun2012,BBRW2021}%
. Indeed, recently
\citet{SNCM2023}%
\ capture covert networks with TUX games to evaluate agents' importance in
such networks, demonstrating that TUX games can lead to more powerful models
and insights than TU games. Whereas the Shapley value probably is the
most-prominent single-valued solution concept for TU\ games, there is a lack
of clarity when it comes to solutions for TUX games. Indeed, many
\textit{generalizations of the Shapley value\emph{\/}} to TUX games were
proposed in the literature, usually motivated by characterizations in the
spirit of Shapley's original characterization based on additivity.\footnote{%
\citet{myerson1977pffg}%
,
\citet{bolger1989}%
,
\citet{AlArRu2005}%
,
\citet{PhaNor2007}%
,
\citet{MSPCWe2007}%
,
\citet{mcquillin2009}%
,
\citet{DuEhKa2010}%
,
\citet{GraFun2012}%
, and
\citet{SkMiWo2018}
introduce different (classes of) solutions.} However, the Shapley value (and
derivatives thereof) are often justified by axioms that involve varying player
sets, such as the balanced contributions property
\citep{myerson1980}%
\ and consistency properties deriving from
\citet{HarMas1989}%
\ and from
\citet{sobolev1975}%
. Therefore,
\citet{DuEhKa2010}%
\ and
\citet{CaFuHu-idd}%
\ argue that generalizations of the Shapley value should also inherit its
properties that relate to varying player sets. After all, these properties not
only lie at the heart of many fairness properties of the Shapley value, but
also build the fundament for implementations via non-cooperative games (see,
e.g., {\citet{gul1989}}, {\citet{MSPCWe2007}}, {\citet{McQSug2016}},
{\citet{BruGauMen2018}}).

\subsection{Novel axioms for games with externalities and characterizations of
the MPW Solution}

In this paper, we define a notion of \emph{balanced contributions for games
with externalities}\textit{\emph{\/}} along the lines of
\citet{myerson1980}%
, which requires that the impact of removing player $j$ on player $i$'s payoff
is the same as the impact of removing player $i$ on player $j$'s payoff. We
demonstrate that this property together with efficiency is characteristic of
the solution introduced by
\citet{feldman1996}%
\ and characterized by
\citet*{MSPCWe2007}%
, henceforth abbreviated as \emph{MPW~solution}, a generalization of the
Shapley value (Theorem~\ref{thm:BCx+EFx<=>sh-star}).

Moreover, we investigate the consistency properties introduced by
\citet{HarMas1989}%
\ and by
\citet{sobolev1975}%
, which are well-known to be characteristic of the Shapley value together with
efficiency and standardness for two-player games. We \emph{augment these
consistency properties to TUX\ games}. Our second main contribution are two
novel characterizations of the MPW\ solution
(Theorems~\ref{thm:sob+e+2s=sh-star} and~\ref{thm:HMC-star+2s=sh-star}).
Generalizing
\citet{sobolev1975}%
\ to games with externalities, we show that the generalized version of
Sobolev's consistency together with efficiency and standardness for two-player
games is characteristic of the MPW~solution.\emph{ }Generalizing
\citet{HarMas1989}%
\ to games with externalities, we show that the generalized version of Hart
and Mas-Colell's consistency together with standardness for two-player games
is characteristic of the MPW~solution. In this sense, we contribute to the
stream of literature that generalizes these consistency properties to broader
frameworks with the intend to generalize the Shapley value (see
\citet{winter1992}%
,
\citet{DuEhKa2010}%
, or
\citet{XuDrSuSu2013}%
\ for other such attempts).

\subsection{Relation to the Literature}

Several motivations for studying balanced contributions and consistency arise
from the literature. We will give a brief summary.

\subsubsection{Balanced contributions property is a fairness property that is
fruitful for derived axiomatizations}

Unlike additivity, the balanced contributions property is a fairness
statement
\citep{moulin2003}%
. It establishes an equitable relationship between players' contributions to
each other. This property has proven instrumental in characterizing various
extensions of the Shapley value (e.g.,
\citet{LFAMCMHe2007}%
,
\citet{GRVP2010}%
,
\citet{KamKon2012}%
, and
\citet{BrGAMaPo2014}%
). The balanced contributions property was also key for motivating specific
applications of the Shapley value, including less traditional applications in
machine learning
\citep{DPGJCM2022}%
, as well as applications in more traditional realms. Recent examples involve
network games
\citep{GAMaPo2015}%
, revenue sharing
\citep{BerMT2023}%
, and group decision models
\citep{MGPC2023}%
. These applications demonstrate how balanced contributions can effectively
embody core fairness principles when adapted to particular operational
contexts. It further suggests that extending this axiom to games with
externalities can lead to a plethora of derivative insights.

\subsubsection{Merits of consistency properties}

Exploring the consistency properties within cooperative game theory serves a
tripartite purpose
\citep{driessen1991}%
. First, it enables a clear differentiation between various solution concepts,
offering a nuanced perspective on their unique properties and conditions of
application. For comparison,
\citet{AMoEhl-prenuk}%
\ provide a characterization of the prenucleolus for TUX games based on a
consistency property generalizing the notion of the reduced game introduced
by
\citet{DavMas1965}%
. Second, exploring consistency properties fosters the theoretical evolution
of these solution concepts, deepening our understanding of their underpinnings
and potential for refinement. This is, for example, important for the
development of mechanisms that implement a solution concept by a
non-cooperative game. To this end, it is particularly useful to know about
ways to pay out players and reduce the size or complexity of a game without
changing the payoffs of the remaining players. Prominent examples for
mechanisms that implement the Shapley value and that stepwise reduce the
number of players are given by
\citet{MSPCWe2007}%
,
\citet{McQSug2016}%
, and
\citet{McQSug2018}%
; implementations of derivatives of the Shapley value along such lines are
given by, for example,
\citet{BeVP2010}%
,
\citet{JuChBr2014}%
, and
\citet{BeGrSa2023}%
. Lastly, the consistency property may be useful in order to determine
coherent solutions for realistic problems, that is, when the game derives from
specific applications. Important examples for which consistency properties of
TU\ games boil down to meaningful properties for the specific application
include airport problems
\citep{LitOwe1973,PotSud1999}%
; bankruptcy problems
\citep{oneill1982,aumann_maschler1985,thomson2015}%
; sequencing problems
\citep{CuPeTi1989,BriChu2011}%
; highway problems
\citep{KuMoZa2013,SudZar2017}%
; queuing problems
\citep{maniquet2003,BenHav2018,ThoVel2022}%
; minimal cost spanning tree problems
\citep{DutKar2004}%
; probabilistic assignment
\citep{chambers2004}%
; and allocating greenhouse gas emission costs
\citep{GoGrGr2021}%
. Even though such allocation problems mostly lend themselves to be studied
with externalities, this has largely been committed from the analysis so far.
Our work introduces a tool that incorporates both externalities and
consistency principles for a general model of allocation problems, which
allows for a reexamination of these models that includes externalities.

\subsubsection{Characterizations of the MPW\ solution\label{subsec:introMPW}}

The MPW solution was first introduced by
\citet{feldman1996}%
\ and characterized by
\citet{MSPCWe2007}%
\ who use Shapley's classical axioms of linearity, efficiency, and the dummy
player property in combination with a strengthening of the symmetry property
and a similar influence requirement.
\citet{fujinaka2004}%
\ provides characterizations of solutions---among others of the $\mathrm{MPW}%
$\ solution---in the spirit of
\citet{young1985}%
, ensuring that each player's reward depends only on this player's vector of
average marginal contributions.
\citet{SkMiWo2018}%
\ highlight the relationship of the Ewens distribution to the
\textquotedblleft Chinese restaurant process\textquotedblright\
(\citealp[11.19]{aldous1985}; \citealp[Equation~3.3]{pitman2006})%
\ and that the $\mathrm{MPW}$\ solution emerges as the expected marginal
contribution of this stochastic coalition formation process.
\citet{SkiMic2020}%
\ deduce a characterization based on Shapley's classical axioms and on
properties of the probability distributions employed in the stochastic
coalition formation process.%

\citet{CaFuHu-idd}%
\ argue that the $\mathrm{MPW}$\ solution is the only plausible generalization
of the\ Shapley value that---like the Shapley value---admits a potential
\citep{HarMas1989}%
, that in turn can be obtained as an expected accumulated worth of partitions
\citep{casajus-potential}%
. We contribute to this literature by providing novel characterizations based
on equal-gains and consistency principles.

\subsection{Structure the paper}

The remainder of this paper is structured as follows. In Section~\ref{sec:def}%
, we introduce basic definitions for TU\ games and TUX games. In
Section~\ref{sec:bc}, we revisit the balanced contributions property
characteristic of the Shapley value, augment balanced contributions to TUX
games, and use it for a characterization of the MPW solution. In
Section~\ref{sec:cons}, we revisit the consistency properties characteristic
of the Shapley value, generalize these consistency properties to TUX\ games,
and use them for novel characterizations of the MPW solution. We end with some
concluding remarks. The appendix contains all the proofs.

\section{Basic definitions and notation\label{sec:def}}

Let $\mathbf{U}$ be a finite set of players, the \textbf{universe of players}.
For $N\subseteq\mathbf{U},$ let $2^{N}$ denote the set of all subsets of
$N.$\footnote{\label{fn:finite-ok-no-players-added}Note that we do not add
players in the proofs, so that a finite universe of players is sufficient.}
Throughout the paper, the cardinalities of coalitions $N,S,T,B\subseteq
\mathbf{U}$ are denoted by $n,$ $s,$ $t,$ and $b,$ respectively.

\subsection{Games without externalities and the Shapley value}

A cooperative game with transferable utility, henceforth \textbf{TU game}
(also known as a \textbf{game in characteristic function form}), for a player
set $N\subseteq\mathbf{U}$ is given by its \textbf{characteristic function}
$v:2^{N}\rightarrow\mathbb{R}$, $v\left(  \emptyset\right)  =0,$ which assigns
a worth to each coalition $S\subseteq N$. Let $\mathbb{V}\left(  N\right)
\ $denote the set of all TU games for $N$ and let $\mathbb{V}$ denote the set
of all TU games. For $v\in\mathbb{V}\left(  N\right)  ,$ $N\subseteq
\mathbf{U}$, and $T\subseteq N,$ the \textbf{restriction} of $v$ by
\textbf{removing the players} in $T$, $v_{-T}\in\mathbb{V}\left(  N\setminus
T\right)  ,$ is given by $v_{-T}\left(  S\right)  =v\left(  S\right)  $ for
all $S\subseteq N\setminus T.$ Alternatively, one can address the game
$v_{-T}$ as the \textbf{subgame} of $v$ on the player set $N\setminus T.$

A \textbf{solution for TU\ games} is an operator $\varphi\mathbf{\ }$that
assigns a payoff vector $\varphi\left(  v\right)  \in\mathbb{R}^{N}$ to any TU
game $v\in\mathbb{V}\left(  N\right)  ,$ $N\subseteq\mathbf{U}.$ The
\textbf{Shapley value }%
\citep{shapley1953}%
, $\mathrm{Sh}$,\ is given by%
\begin{equation}
\mathrm{Sh}_{i}\left(  v\right)  =\sum_{S\subseteq N\setminus\left\{
i\right\}  }\frac{s!\left(  n-s-1\right)  !}{n!}\left(  v\left(  S\cup\left\{
i\right\}  \right)  -v\left(  S\right)  \right)  \label{eq:Sh}%
\end{equation}
for all $v\in\mathbb{V}\left(  N\right)  ,$ $N\subseteq\mathbf{U}$ and $i\in
N.$

\subsection{Games with externalities and the MPW solution}

A \textbf{partition} of $N\subseteq\mathbf{U}$ is a collection of non-empty
subsets of $N$ such that any two of them are disjoint and such their union is
$N.$ The set of partitions of $N\subseteq\mathbf{U}$ is denoted by $\Pi\left(
N\right)  .$ For technical reasons, we set $\Pi\left(  \emptyset\right)
=\left\{  \emptyset\right\}  .$ The block of $\pi\in\Pi\left(  N\right)  $
that contains player $i\in N$ is denoted by $\pi\left(  i\right)  $.

For $\pi\in\Pi\left(  N\right)  ,$ $N\subseteq\mathbf{U},$ the
\textbf{elimination of the players} in $T\subseteq N$ from $\pi$ gives
$\pi_{-T}\in\Pi\left(  N\setminus T\right)  ,$%
\[
\pi_{-T}=\left\{  \left\{  B\setminus T\right\}  \mid B\in\pi\text{ and
}B\setminus T\neq\emptyset\right\}  .
\]
Instead of $\pi_{-\left\{  i\right\}  }$, we write $\pi_{-i}.$ For
$N\subseteq\mathbf{U}$ and $\pi\in\Pi\left(  N\right)  ,$ \textbf{adding a
player} $i\in\mathbf{U}\setminus N$ to the block $B\in\pi$ is denoted by
$\pi_{+i\rightsquigarrow B}\in\Pi\left(  N\cup\left\{  i\right\}  \right)  ,$
\[
\pi_{+i\rightsquigarrow B}=\left(  \pi\setminus\left\{  B\right\}  \right)
\cup\left\{  B\cup\left\{  i\right\}  \right\}  ;
\]
adding player $i$ as a singleton is denoted by $\pi_{+i\rightsquigarrow
\emptyset}\in\Pi\left(  N\cup\left\{  i\right\}  \right)  ,$ $\pi
_{+i\rightsquigarrow\emptyset}=\pi\cup\left\{  \left\{  i\right\}  \right\}  $.

A TU game with externalities, henceforth \textbf{TUX game }(also known as
\textbf{game in partition function form}), for a player set $N\subseteq
\mathbf{U}$ is given by its \textbf{partition function} $w:\mathcal{E}\left(
N\right)  \rightarrow\mathbb{R}^{N},$ where $\mathcal{E}\left(  N\right)  $
denotes the set of \textbf{embedded coalitions} $\left(  S,\pi\right)  $ for
$N$ given by
\[
\mathcal{E}\left(  N\right)  =\left\{  \left(  S,\pi\right)  \mid S\subseteq
N\text{ and }\pi\in\Pi\left(  N\setminus S\right)  \right\}  ,
\]
and with $w\left(  \emptyset,\pi\right)  =0\ $for$\ $all$\ \pi\in\Pi\left(
N\right)  $. We denote the set of all TUX games for a player set $N$ by
$\mathbb{W}\left(  N\right)  $ and the set of all TUX games by\ $\mathbb{W}$.
For $N\subseteq\mathbf{U},$ the \textbf{null game} $\mathbf{0}^{N}%
\in\mathbb{W}\left(  N\right)  $ is defined by $\mathbf{0}^{N}\left(
S,\pi\right)  =0$ for all $\left(  S,\pi\right)  \in\mathcal{E}\left(
N\right)  .$

\subsubsection{The MPW solution\label{sec:MPW}}

Several solutions were introduced in the literature to generalize the Shapley
value, i.e., solutions that boil down to the Shapley value if a TUX\ game
actually does not exhibit externalities, i.e., if $w\left(  S,\pi\right)
=w\left(  S,\tau\right)  $ for all $S\subseteq N$ and $\pi,\tau\in\Pi\left(
N\setminus S\right)  $.
\citet{MSPCWe2007}%
\ put forth a solution for TUX games, $\mathrm{MPW},$ following a two step
procedure:\footnote{The MPW\ solution is part of a larger class
called\textit{\emph{\/}} average Shapley values introduced and characterized
by
\citet{MSPCWe2007}%
, which includes solutions derived by an analogous two step procedure but with
possibly other probability distributions over $\Pi\left(  N\setminus S\right)
$.}

\begin{enumerate}
\item For a given TUX game $w\in\mathbb{W}\left(  N\right)  ,$ $N\subseteq
\mathbf{U}$, one first computes a~TU\ game $\bar{v}_{w}\in\mathbb{V}\left(
N\right)  $, the average game, in which each coalition~$S$ gets the following
expected value of $w\left(  S,\pi\right)  $ over all partitions $\pi\in
\Pi\left(  N\setminus S\right)  $,%
\begin{equation}
\bar{v}_{w}\left(  S\right)  =\sum_{\pi\in\Pi\left(  N\setminus S\right)
}\frac{\prod_{B\in\pi}\left(  b-1\right)  !}{\left(  n-s\right)  !}w\left(
S,\pi\right)  \qquad\text{for all }S\subseteq N. \label{eq:averagegame}%
\end{equation}

\item Second, one applies the Shapley value to this TU game in order to obtain
the MPW solution,
\begin{equation}
\mathrm{MPW}\left(  w\right)  =\mathrm{Sh}\left(  \bar{v}_{w}\right)
\qquad\text{for all }w\in\mathbb{W}\left(  N\right)  ,~N\subseteq\mathbf{U}.
\label{eq:Sh-star}%
\end{equation}

\end{enumerate}

Note that
\begin{equation}
p_{N\setminus S}^{\star}\left(  \pi\right)  =\prod_{B\in\pi}\frac{\left(
b-1\right)  !}{\left(  n-s\right)  !} \label{eq:p-star}%
\end{equation}
for all $S\subseteq N\subseteq\mathbf{U}$ and $\pi\in\Pi\left(  N\setminus
S\right)  $ is a probability distribution over the set of partitions
$\Pi\left(  N\setminus S\right)  $. A family of probability distributions
$p=\left(  p_{N}\right)  _{N\subseteq\mathbf{U}}$ over partitions is called a
random partition\emph{ }for $\mathbf{U}$. The random partition $p^{\star}$ is
known as the Ewens distribution with mutation rate\emph{ }$\theta=1$
\citep{ewens1972}%
, which takes a central role in the literature on random partitions
\citep{crane2016}%
.

\subsubsection{Subgames for TUX games\label{sec:subgamesTUX}}

For TU games, there is an obvious way to obtain subgames. In contrast, the
notion of a subgame is less obvious for TUX games, since we cannot simply read
it off the original game. When player $i$ is removed from the TUX game $w$, we
have to specify the worth of each embedded coalition \textquotedblleft%
$w_{-i}\left(  S,\pi\right)  $\textquotedblright\ in the TUX game $w_{-i}$
without player $i$. For instance, when removing player $4$ from some TUX game
$w\in\mathbb{W}\left(  \left\{  1,2,3,4\right\}  \right)  $, the worth
$w_{-4}\left(  \left\{  1\right\}  ,\left\{  \left\{  2,3\right\}  \right\}
\right)  $ has no obvious reference in the original game $w$, where player $4$
impacts the worth of coalition~$\left\{  1\right\}  $ through being singleton
or being affiliated with $\left\{  2,3\right\}  $.

To capture the many possibilities of how to obtain subgames in the presence of
externalities,
\citet{DuEhKa2010}
introduce the concept of a \textbf{restriction operator}. A restriction
operator $r$ formally specifies how to obtain a \textquotedblleft
subgame\textquotedblright\ $w_{-i}^{r}\in\mathbb{W}\left(  N\setminus\left\{
i\right\}  \right)  $ for every TUX game $w\in\mathbb{W}\left(  N\right)  ,$
$N\subseteq\mathbf{U},$ and every player $i\in N$, where the worth $w_{-i}%
^{r}\left(  S,\pi\right)  $ in the subgame only depends on the worths
$w\left(  S,\pi_{+i\leadsto\emptyset}\right)  $ and $w\left(  S,\pi
_{+i\rightsquigarrow B}\right)  $ for $B\in\pi$ in the original game $w$. For
example, a restriction operator $r$ specifies how the worth $w_{-4}^{r}\left(
\left\{  1\right\}  ,\left\{  \left\{  2,3\right\}  \right\}  \right)  $ is
derived from aggregating the numbers $w\left(  \left\{  1\right\}  ,\left\{
\left\{  2,3\right\}  ,\left\{  4\right\}  \right\}  \right)  $ and $w\left(
\left\{  1\right\}  ,\left\{  \left\{  2,3,4\right\}  \right\}  \right)  $.
Subgames that are constructed in this way can be seen as \textquotedblleft%
`estimates'\ or `approximations'\ based on the available
data\textquotedblright\
\citep{DuEhKa2010}%
.

To ensure that subgames are well-defined even if more than one player was
removed, i.e., to consider $w_{-T}^{r},$ $T\subseteq N,$ it must not matter in
which order the players are removed, $(w_{-i}^{r})_{-j}^{r}=(w_{-j}^{r}%
)_{-i}^{r}$ for all $N\subseteq\mathbf{U},$ $w\in\mathbb{W}\left(  N\right)
,$ and $i,j\in N,$ $i\neq j$. A restriction operator satisfying this principle
is said to be path independent.

Whilst there exists a plethora of possible path independent restrictions
operators,
\citet{CaFuHu-idd}
argue there is a unique way to define a restriction operator that maintains
various properties of the Shapley value.

\subsubsection{The restriction operator $r^{\star}$\label{subsec:r-star}}%

\citet{HarMas1989}
show that the Shapley value is a player's contribution to the so-called
potential of a game, which can be obtained as the expected accumulated worth
under a probabilistic partitioning scheme
\citep{casajus-potential}%
. Specifically, given a suitable probability distribution defined over the set
of all possible partitions of the player set, the potential function equals
the expected value of the sum of coalition worths within each randomly
selected partition. Imposing that a potential for TUX\ games maintains this
interpretation, and resting on the fact every path-independent restriction
operator corresponds to a potential for TUX\ games
\citep[Theorem~1]{DuEhKa2010}%
,
\citet{CaFuHu-idd}
single out a unique restriction operator $r^{\star}$.\footnote{%
\citet{CaFuHu-idd}%
\ impose further assumptions to ensure that the obtained potential for TUX
games generalizes the potential for TU\ games.}

In what follows, we will limit the analysis to the restriction operator
$r^{\star}$, and we will omit explicit mention of the superscript. It is
defined by%
\begin{equation}
w_{-i}\left(  S,\pi\right)  =w_{-i}^{r^{\star}}\left(  S,\pi\right)  =\frac
{1}{n-s}\left(  w\left(  S,\pi_{+i\rightsquigarrow\emptyset}\right)
+\sum_{j\in N\setminus\left(  S\cup\left\{  i\right\}  \right)  }w\left(
S,\pi_{+i\rightsquigarrow\pi\left(  j\right)  }\right)  \right)
\label{eq:r-star}%
\end{equation}
for all $w\in\mathbb{W}\left(  N\right)  $, $N\subseteq\mathbf{U}$, $i\in N,$
and $\left(  S,\pi\right)  \in\mathcal{E}\left(  N\setminus\left\{  i\right\}
\right)  $. Thus, to compute worth of an embedded coalition $\left(
S,\pi\right)  $ without player $i$, we perform a counterfactual analysis. This
analysis considers what would happen if player $i$ were not explicitly part of
coalition $S$. To do this, we imagine player $i$ engaging in all possible
interactions with players outside of $S$, or choosing to remain isolated.
Giving equal chance to player $i$ making connections with any individual
player $j\in N\setminus S$ who is not in coalition $S$ (making a connection
with player $i$ themselves is interpreted as staying alone), we expect the
worth $w_{-i}\left(  S,\pi\right)  $. This simple average over the worths of
embedded coalitions then reflects the potential externalities due to player
$i$. For example, removing player~$4$ from $w\in\mathbb{W}\left(  \left\{
1,2,3,4\right\}  \right)  $ gives
\[
w_{-4}^{r}\left(  \left\{  1\right\}  ,\left\{  \left\{  2,3\right\}
\right\}  \right)  =\frac{1}{3}w\left(  \left\{  1\right\}  ,\left\{  \left\{
2,3\right\}  ,\left\{  4\right\}  \right\}  \right)  +\frac{2}{3}w\left(
\left\{  1\right\}  ,\left\{  \left\{  2,3,4\right\}  \right\}  \right)  .
\]

It is important to note that a removed player casts a shadow on the subgame.
For example, a subgame originating from a four-player Cournot oligopoly is
distinct from a three-firm Cournot oligopoly.

\begin{example}
Consider a Cournot oligopoly with four firms $N=\left\{  1,2,3,4\right\}  $,
which have identical constant marginal cost $c>0$, and which face inverse
demand $P\left(  X\right)  =A-X_{N}$ with $A>c,$ where the output of some
coalition $S\subseteq N$ is given by $X_{S}=\sum_{i\in S}x_{i}$. Given a
partition $\pi\in\Pi\left(  N\right)  $ into $\left\vert \pi\right\vert $
cartels, each cartel $S\in\pi$ chooses the joint quantity $X_{S}$ to maximizes
its joint profit $\left(  A-c-X_{N\setminus S}-X_{S}\right)  X_{S}$. In the
Cournot equilibrium, each cartel $S\in\pi$ then has the same profit%
\[
w\left(  S,\pi\setminus\left\{  S\right\}  \right)  =\frac{\left(  A-c\right)
^{2}}{\left(  \left\vert \pi\right\vert +1\right)  ^{2}}.
\]
Note that profits are larger the less cartels there are, i.e., merging cartels
exercise positive external effects on the other cartels (see
\citet{yi1997}%
\ for a more precise definition of positive/negative external effects in this context).
\end{example}

Reducing this TUX game to a subgame now means capturing the strategic
interactions by a TUX game with the remaining firms. The possibility of
external effects of the removed firm is still accounted for, but the removed
firm's ability to actively participate in coalition formation with the
remaining firms is no longer considered. Instead, when modeling the profit of
a cartel in the subgame, we take a probabilistic approach concerning the
alliance of the removed firm with outside firms. The restriction without firm
$4$, $w_{-4}\in\mathbb{W}\left(  \left\{  1,2,3\right\}  \right)  $ is given
by
\begin{align*}
w_{-4}\left(  \left\{  i\right\}  ,\left\{  \left\{  j\right\}  ,\left\{
k\right\}  \right\}  \right)   &  =\frac{1}{3}w\left(  \left\{  i\right\}
,\left\{  \left\{  j\right\}  ,\left\{  k\right\}  ,\left\{  4\right\}
\right\}  \right)  +\frac{2}{3}w\left(  \left\{  i\right\}  ,\left\{  \left\{
j,4\right\}  ,\left\{  k\right\}  \right\}  \right)  =\frac{11}{200}\left(
A-c\right)  ^{2}\\
w_{-4}\left(  \left\{  i\right\}  ,\left\{  \left\{  j,k\right\}  \right\}
\right)   &  =\frac{1}{3}w\left(  \left\{  i\right\}  ,\left\{  \left\{
j,k\right\}  ,\left\{  4\right\}  \right\}  \right)  +\frac{2}{3}w\left(
\left\{  i\right\}  ,\left\{  \left\{  j,k,4\right\}  \right\}  \right)
=\frac{41}{432}\left(  A-c\right)  ^{2}\\
w_{-4}\left(  \left\{  i,j\right\}  ,\left\{  \left\{  k\right\}  \right\}
\right)   &  =\frac{1}{2}w\left(  \left\{  i,j\right\}  ,\left\{  \left\{
k\right\}  ,\left\{  4\right\}  \right\}  \right)  +\frac{1}{2}w\left(
\left\{  i,j\right\}  ,\left\{  \left\{  k,4\right\}  \right\}  \right)
=\frac{25}{288}\left(  A-c\right)  ^{2}\\
w_{-4}\left(  \left\{  1,2,3\right\}  ,\left\{  \emptyset\right\}  \right)
&  =w\left(  \left\{  1,2,3\right\}  ,\left\{  \left\{  4\right\}  \right\}
\right)  =\frac{1}{9}\left(  A-c\right)  ^{2}%
\end{align*}
for $\left\{  i,j,k\right\}  =\left\{  1,2,3\right\}  $. We observe that
$w_{-4}\left(  \left\{  i\right\}  ,\left\{  \left\{  j,k\right\}  \right\}
\right)  >w_{-4}\left(  \left\{  i,j\right\}  ,\left\{  \left\{  k\right\}
\right\}  \right)  $. This difference arises from the higher probability that
firm $4$ is believed to connect with $\left\{  j,k\right\}  $ compared to
$\left\{  k\right\}  $. Consequently, the likelihood of firm 4 operating
independently (and thus not contributing positive externalities to the
embedded coalition) is reduced. Clearly, this subgame differs from a standard
three-firm Cournot oligopoly, $N=\left\{  1,2,3\right\}  $. For instance,
\[
w\left(  \left\{  1,2,3\right\}  \right)  =\frac{1}{4}\left(  A-c\right)
^{2}>\frac{1}{9}\left(  A-c\right)  ^{2}=w_{-4}\left(  \left\{  1,2,3\right\}
,\left\{  \emptyset\right\}  \right)  .
\]
The subgame $w_{-4}$ acknowledges the negative externalities exerted by firm 4
on the profit of the grand coalition $\left\{  1,2,3\right\}  $. Therefore,
while this restriction accounts for the removed firm's externalities, it
simultaneously constrains any potential cartel formation with that firm.

\section{Balanced contributions \label{sec:bc}}

We start by revisiting the definition of balanced contributions for TU\ games
and the characterization of the Shapley value based on this property.
Thereafter, we extend the result to TUX\ games.

\subsection{Myerson's characterization of the Shapley value}%

\citet{myerson1980}%
\ observed that the Shapley value satisfies the following property.

\medskip

\noindent\emph{Balanced Contributions}\textbf{, BC.}$\;$For all $v\in
\mathbb{V}\left(  N\right)  ,~N\subseteq\mathbf{U},$ and $i,j\in N$, we have
$\varphi_{i}\left(  v\right)  -\varphi_{i}\left(  v_{-j}\right)  =\varphi
_{j}\left(  v\right)  -\varphi_{j}\left(  v_{-i}\right)  $.

\medskip

According to this requirement the impact of removing player $j$ on player
$i$'s payoff is the same as the impact of removing player $i$ on player $j$'s
payoff. There is a rich literature investigating the consequences of this
property. Equivalent properties are summarized by
\citet{CasHue-deva}%
. Not much is missing to obtain a characterization of the Shapley value.

\medskip

\noindent\emph{Efficiency, }\textbf{EF.}$\;$For all $v\in\mathbb{V}\left(
N\right)  ,~N\subseteq\mathbf{U},$ we have $\sum_{i\in N}\varphi_{i}\left(
v\right)  =v\left(  N\right)  .$

\medskip

Perhaps surprisingly, balanced contributions together with efficiency already
is characteristic of the Shapley value.

\begin{theorem}
[Myerson, 1980]The Shapley value, $\mathrm{Sh}$, is the unique solution for TU
games that satisfies efficiency (\textbf{EF}) and the balanced contributions
property (\textbf{BC}).
\end{theorem}

Next, we turn to the case with externalities.

\subsection{A characterization of the MPW solution based on balanced
contributions}

Based on the restriction defined in (\ref{eq:r-star}), we obtain a
straight-forward generalization of the balanced contributions property to TUX games.

\medskip

\noindent\emph{Balanced Contributions}\textbf{, BC}$^{\text{\textbf{X}}}%
$\textbf{.}$\;$For all $w\in\mathbb{W}\left(  N\right)  ,~N\subseteq
\mathbf{U},$ and $i,j\in N$, we have $\varphi_{i}\left(  w\right)
-\varphi_{i}\left(  w_{-j}\right)  =\varphi_{j}\left(  w\right)  -\varphi
_{j}\left(  w_{-i}\right)  $.

\medskip

The impact of removing player $j$ on player $i$'s payoff is the same as the
impact of removing player $i$ on player $j$'s payoff, when removing a player
is accounted for in the fashion proposed by
\citet{CaFuHu-idd}%
. Consider the following example introduced by
\citet{maskin2003}%
\ to get a better intuition for this property.

\begin{example}
\label{ex:maskin2003}Let a three-player public goods game be given by
$N=\left\{  1,2,3\right\}  $ and
\begin{align*}
w\left(  \left\{  i\right\}  ,\left\{  \left\{  j\right\}  ,\left\{
k\right\}  \right\}  \right)   &  =0;\\
w\left(  \left\{  i\right\}  ,\left\{  \left\{  j,k\right\}  \right\}
\right)   &  =9;\\
w\left(  \left\{  1,2\right\}  ,\left\{  \left\{  3\right\}  \right\}
\right)   &  =12;\\
w\left(  \left\{  1,3\right\}  ,\left\{  \left\{  2\right\}  \right\}
\right)   &  =13;\\
w\left(  \left\{  2,3\right\}  ,\left\{  \left\{  1\right\}  \right\}
\right)   &  =14;\\
w\left(  \left\{  1,2,3\right\}  ,\left\{  \emptyset\right\}  \right)   &
=24.
\end{align*}

\end{example}

The subgames $w_{-i}$, $i\in\left\{  1,2,3\right\}  $ reflect that an
affiliation of the removed player with outside coalitions induces
externalities,
\[
w_{-i}\left(  \left\{  j\right\}  ,\left\{  \left\{  k\right\}  \right\}
\right)  =\frac{1}{2}w\left(  \left\{  i\right\}  ,\left\{  \left\{
j\right\}  ,\left\{  k\right\}  \right\}  \right)  +\frac{1}{2}w\left(
\left\{  i\right\}  ,\left\{  \left\{  j,k\right\}  \right\}  \right)  =4.5.
\]
The subgames also reflect the asymmetries between the players: $w_{-1}\left(
\left\{  23\right\}  ,\left\{  \emptyset\right\}  \right)  =14$,
$w_{-2}\left(  \left\{  13\right\}  ,\left\{  \emptyset\right\}  \right)
=13$, and $w_{-3}\left(  \left\{  12\right\}  ,\left\{  \emptyset\right\}
\right)  =12$.

TUX\ games with two players are in fact TU\ games. Applying the Shapley value
to the restricted games gives the payoffs $\varphi_{1}\left(  w_{-2}\right)
=6.5$, $\varphi_{1}\left(  w_{-3}\right)  =6$, $\varphi_{2}\left(
w_{-1}\right)  =7$, and $\varphi_{3}\left(  w_{-1}\right)  =7$. Balanced
contributions addresses these asymmetries and requires equal gains,
\begin{align*}
\varphi_{1}\left(  w\right)  -6.5  &  =\varphi_{2}\left(  w\right)  -7;\\
\varphi_{1}\left(  w\right)  -6  &  =\varphi_{3}\left(  w\right)  -7.
\end{align*}
Note that together with efficiency, $\varphi_{1}\left(  w\right)  +\varphi
_{2}\left(  w\right)  +\varphi_{3}\left(  w\right)  =24$, we have enough
independent equations to infer $\varphi_{1}\left(  w\right)  =7.5$,
$\varphi_{2}\left(  w\right)  =8$, and $\varphi_{3}\left(  w\right)  =8.5$.

In this example, the payoffs coincide with the efficient generalized Shapley
value introduced by
\citet{hafalir2007}%
. In contrast, the payoffs according to the \textquotedblleft
externality-free\ value\textquotedblright\
\citep{CliSer2008}%
, which was\ introduced by
\citet{PhaNor2007}%
, follow from the assumption that outside players always stay singletons
(hence, it ignores the data $w\left(  \left\{  i\right\}  ,\left\{  \left\{
j,k\right\}  \right\}  \right)  =9$) and are given by $\left(  \frac{43}%
{6},\frac{44}{6},\frac{45}{6}\right)  $.

It turns out that the balanced contributions property together with efficiency
is characteristic of the MPW solution. We prepare the statement of this result
with a formal definition of efficiency for TUX\ games.

\medskip

\noindent\emph{Efficiency, }\textbf{EF}$^{\text{\textbf{X}}}$\textbf{.}$\;$For
all $w\in\mathbb{W}\left(  N\right)  ,~N\subseteq\mathbf{U},$ we have
$\sum_{i\in N}\varphi_{i}\left(  w\right)  =w\left(  N,\emptyset\right)  .$

\medskip

We can now state our first main result.

\begin{theorem}
\label{thm:BCx+EFx<=>sh-star}The MPW solution, $\mathrm{MPW}$, is the unique
solution for TUX games that satisfies efficiency (\textbf{EF}%
$^{\text{\textbf{X}}}$) and the balanced contributions property (\textbf{BC}%
$^{\text{\textbf{X}}}$).
\end{theorem}

Whereas the proof of uniqueness rests on an inductive argument, it may be less
obvious that the MPW\ solution actually satisfies the balanced contributions
property. However, this is a consequence of the fact that the Shapley value
satisfies the balanced contributions property, the fact that the MPW solution
is the Shapley value of the average game, and the following lemma, which
states that it does not matter if we first average a game and then remove a
player from the obtained TU game, $\left(  \bar{v}_{w}\right)  _{-i}$; or
whether we first remove a player from the TUX game and then take the average
second $\bar{v}_{\left(  w_{-i}\right)  }$.

\begin{lemma}
\label{lem:r-and-avg-commute}For all $w\in\mathbb{W}\left(  N\right)  $,
$N\subseteq\mathbf{U}$, and $i\in N$, we have $\left(  \bar{v}_{w}\right)
_{-i}=\bar{v}_{\left(  w_{-i}\right)  }\in\mathbb{V}\left(  N\setminus\left\{
i\right\}  \right)  $.
\end{lemma}

It turns out that this commutative relationship of removal and average
operators is instrumental for further results as demonstrated in the next section.

\section{Consistency\label{sec:cons}}

An important stream of characterization results of the Shapley value and
alternative solution concepts for TU games relies on consistency properties.
These properties involve the notion of a reduced game, which is formed when
one or more players are removed from the original game, with the understanding
that these players receive compensation as determined by a specific payoff
principle. The reduced game thus depends on the original game, the payoffs
allocated to the removed players, and (sometimes) it also depends on the
solution proposed for a subgame. The consistency property asserts that if a
payoff vector exists for the original game, then a corresponding payoff vector
should be achievable for the reduced game's players---ensuring a seamless
transition in value distribution despite changes in player configuration.

For the characterization of the Shapley value,
\citet{PelSud2007}%
\ highlight the consistency properties introduced by
\citet{sobolev1975}%
\ and
\citet{HarMas1989}%
, which we revisit next. Thereafter, we introduce consistency properties that
generalize these properties to TUX games.

\subsection{Sobolev's and Hart and Mas-Colell's characterization of the
Shapley value\label{sec:consTU}}

Among the consistency properties for TU games (without externalities)
discussed in the literature, the notion introduced by
\cite{sobolev1975}%
\ has the particular appeal of \emph{not}\textit{\emph{\/}} referring to the
payoffs of subgames; instead, it refers to some sort of a protocol. In this
sense, it appears to capture the nature of the Shapley value particularly well.

Let $\varphi$ be a solution for TU games, and let $v\in\mathbb{V}\left(
N\right)  $, $N\subseteq\mathbf{U}$, and $j\in N$. Sobolev's notion of a
reduced TU\ game $v_{-j}^{\varphi}\in\mathbb{V}(N\setminus\left\{  j\right\}
)$ is given by%
\begin{equation}
v_{-j}^{So,\varphi}\left(  S\right)  =\frac{s}{n-1}\left(  v\left(
S\cup\left\{  j\right\}  \right)  -\varphi_{j}\left(  v\right)  \right)
+\left(  1-\frac{s}{n-1}\right)  v\left(  S\right)  \text{\qquad for all
}S\subseteq N\setminus\left\{  j\right\}  . \label{eq:sobolev-TU-reduced-game}%
\end{equation}
The idea behind this reduced game can be described as follows. Player~$j$ will
join forces with one of the $n-1$ other players, each player equally likely.
The worth of a coalition~$S$ in the reduced game depends on whether player~$j$
joins one of the $s$ players within coalition $S$. If this is the case, then
coalition $S$ makes use of player~$j$'s productivity and compensates
player~$j$ with~$\varphi_{j}\left(  v\right)  $, so that the worth of
coalition $S$ equals $v\left(  S\cup\left\{  j\right\}  \right)  -\varphi
_{j}\left(  v\right)  $; otherwise, player~$j$ joins the other players and the
worth of coalition $S$ remains the same. In expectation, coalition~$S$ has the
worth $v_{-j}^{So,\varphi}\left(  S\right)  $. Interestingly, such a reduction
or amalgamation of player~$j$ is neutral to the Shapley payoff of the other
players, that is, the Shapley value satisfies the following property.

\medskip

\noindent\emph{Sobolev Consistency}\textbf{, SC.}$\;$For all $v\in
\mathbb{V}\left(  N\right)  ,N\subseteq\mathbf{U},$ and $i,j\in N,$ $i\neq j$,
we have $\varphi_{i}\left(  v\right)  =\varphi_{i}(v_{-j}^{So,\varphi})$.

\medskip

It is further useful to compare this with the consistency property proposed
by
\citet{HarMas1989}%
. To this end, define the reduced game $v_{-j}^{HM,\varphi}\in\mathbb{V}%
(N\setminus\left\{  j\right\}  )$ by
\[
v_{-j}^{HM,\varphi}\left(  S\right)  =v\left(  S\cup\left\{  j\right\}
\right)  -\varphi_{j}\left(  v_{-N\setminus\left(  S\cup\left\{  j\right\}
\right)  }\right)  \text{\qquad for all }S\subseteq N\setminus\left\{
j\right\}  .
\]
Here, the worth of coalition $S$ in a subgame is obtained by utilizing the
productivity of the removed player~$j$ and paying to~$j$ the payoff that
player~$j$ obtains in the game without the other players $v_{-N\setminus
\left(  S\cup\left\{  j\right\}  \right)  }$. Again, the Shapley value applied
to the reduced game gives a player the same payoff as in the original game,
that is, the Shapley value satisfies the following property.

\medskip

\noindent\emph{Hart and Mas-Colell Consistency}\textbf{, HMC.}$\;$For all
$v\in\mathbb{V}\left(  N\right)  ,~N\subseteq\mathbf{U},$ and $i,j\in N,$
$i\neq j$, we have $\varphi_{i}\left(  v\right)  =\varphi_{i}(v_{-j}%
^{HM,\varphi})$.

\medskip

\noindent Different from the Sobolev consistency, HM consistency applies the
solution concept to a subgame $v_{-S\setminus T}$ in the definition of the
reduced game.

The Shapley value is the only efficient solution concept that satisfies these
consistency properties and standardness for two-player games, which simply
prescribes that the surplus in a two-player game is equally shared.

\medskip

\noindent$2$\emph{-Standardness, }$\mathbf{2}$\textbf{S.}$\;$For all
$v\in\mathbb{V}\left(  \left\{  i,j\right\}  \right)  ,~\left\{  i,j\right\}
\subseteq\mathbf{U}$, $i\neq j,$ we have
\[
\varphi_{i}\left(  v\right)  =v\left(  \left\{  i\right\}  \right)
+\frac{v\left(  \left\{  i,j\right\}  \right)  -v\left(  \left\{  j\right\}
\right)  -v\left(  \left\{  i\right\}  \right)  }{2}.
\]

The literature emphasizes the following characterizations.

\begin{theorem}
[Sobolev, 1975]The Shapley value, $\mathrm{Sh}$, is the unique solution that
satisfies efficiency (\textbf{EF}), 2-standardness (\textbf{2S}), and Sobolev
consistency (\textbf{SC}).
\end{theorem}

\begin{theorem}
[Hart and Mas-Colell, 1989]\label{thm:HM1989-con}The Shapley value,
$\mathrm{Sh}$, is the unique solution that satisfies 2-standardness
(\textbf{2S}) and Hart and Mas-Colell consistency (\textbf{HMC}).
\end{theorem}

We further remark that
\citet{HarMas1989}
also discuss the following equivalent notion of HM consistency that allows for
the removal of multiple players:

\medskip

\noindent For all $v\in\mathbb{V}\left(  N\right)  ,~N\subseteq\mathbf{U},$
and $T\subseteq N,$ $i\in N\setminus T$, we have
\begin{equation}
\varphi_{i}\left(  v\right)  =\varphi_{i}(v_{-T}^{HM,\varphi}),
\label{eq:setHMC}%
\end{equation}
where $v_{-T}^{HM,\varphi}$ is given
\begin{equation}
v_{-T}^{HM,\varphi}\left(  S\right)  =v\left(  S\cup T\right)  -\sum_{j\in
T}\varphi_{j}\left(  v_{-N\setminus\left(  S\cup T\right)  }\right)
\text{\qquad for all }S\subseteq N\setminus T. \label{eq:setHMreducedgame}%
\end{equation}
Note that for $T=\left\{  j\right\}  $, the reduced game in
(\ref{eq:setHMreducedgame}) becomes $v_{-T}^{HM,\varphi}=v_{-j}^{HM,\varphi},$
and the above property (\ref{eq:setHMC}) boils down to HM consistency; hence,
(\ref{eq:setHMC}) implies HM consistency. Perhaps surprisingly, HM consistency
in turn implies property (\ref{eq:setHMC}) even if $t>1$
\citep[Lemma~4.4]{HarMas1989}%
. This is a key to establish the characterization in
Theorem~\ref{thm:HM1989-con}.

\subsection{Sobolev\emph{ }consistency for TUX Games\label{sec:SobconsTUX}}

When generalizing Sobolev's consistency, we have to be careful with the
formulation of the reduced game. Note that in the definition of the reduced
game, (\ref{eq:sobolev-TU-reduced-game}), the worth $v\left(  S\right)  $
conceptually refers to the worth in the game without player$~j$, that is, to a
subgame. In order to introduce an analogous notion of a reduced game for TUX
games, we need to apply the restriction at this place.

Let $\varphi$ be a solution for TUX games and let $w\in\mathbb{W}\left(
N\right)  $, $N\subseteq\mathbf{U}$, and $j\in N$. The reduced TUX\ game
$w_{-j}^{So,\varphi}\in\mathbb{W}\left(  N\setminus\left\{  j\right\}
\right)  $ is defined by%
\begin{equation}
w_{-j}^{So,\varphi}\left(  S,\pi\right)  =\frac{s}{n-1}\left(  w\left(
S\cup\left\{  j\right\}  ,\pi\right)  -\varphi_{j}\left(  w\right)  \right)
+\left(  1-\frac{s}{n-1}\right)  w_{-j}\left(  S,\pi\right)
\label{eq:sobolev-TUX-reduced-game}%
\end{equation}
for all $\left(  S,\pi\right)  \in\mathcal{E}\left(  N\setminus\left\{
j\right\}  \right)  $, where the subgame $w_{-j}$ is defined in
(\ref{eq:r-star}). This suggests the following consistency property.

\medskip

\noindent\emph{Sobolev Consistency}\textbf{, SC}$^{\text{\textbf{X}}}%
$\textbf{.}$\;$For all $w\in\mathbb{W}\left(  N\right)  ,~N\subseteq
\mathbf{U},$ and $i,j\in N,$ $i\neq j$, we have $\varphi_{i}\left(  w\right)
=\varphi_{i}(w_{-j}^{So,\varphi})$.

\medskip

To illustrate this property, recall the public good game in
Example~\ref{ex:maskin2003}, for which $\mathrm{MPW}_{1}\left(  w\right)
=7.5$ and $w_{-1}\left(  \left\{  2\right\}  ,\left\{  \left\{  3\right\}
\right\}  \right)  =4.5$. The reduced game without player $1$ is given by%
\begin{align*}
w_{-1}^{So,\varphi}\left(  \left\{  2\right\}  ,\left\{  \left\{  3\right\}
\right\}  \right)   &  =\frac{1}{2}\left(  w\left(  \left\{  1,2\right\}
,\left\{  \left\{  3\right\}  \right\}  \right)  -\mathrm{MPW}_{1}\left(
w\right)  \right)  +\frac{1}{2}w_{-1}\left(  \left\{  2\right\}  ,\left\{
\left\{  3\right\}  \right\}  \right)  =4.5;\\
w_{-1}^{So,\varphi}\left(  \left\{  3\right\}  ,\left\{  \left\{  2\right\}
\right\}  \right)   &  =5;\\
w_{-1}^{So,\varphi}\left(  \left\{  23\right\}  ,\left\{  \emptyset\right\}
\right)   &  =w\left(  \left\{  1,2,3\right\}  ,\pi\right)  -\mathrm{MPW}%
_{1}\left(  w\right)  =16.5.
\end{align*}
Indeed, solving this two-player game gives $\mathrm{MPW}_{2}(w_{-1}%
^{So,\varphi})=8$ and $\mathrm{MPW}_{3}(w_{-1}^{So,\varphi})=8.5$, which
coincides with the payoffs in the original game $\mathrm{MPW}_{2}\left(
w\right)  $ and $\mathrm{MPW}_{3}\left(  w\right)  $, respectively.

It is not obvious that the MPW solution satisfies this requirement. We deduce
this from the fact that the Shapley value satisfies Sobolev consistency, the
fact that the MPW solution is the Shapley value on the average game, and
Lemma~\ref{lem:r-and-avg-commute}, which states that it does not matter
whether we first average a game and then remove a player from the obtained TU
game, or whether we first remove a player from to the TUX game $w$ and and
then take the average, i.e., $\left(  \bar{v}_{w}\right)  _{-i}=\bar
{v}_{\left(  w_{-i}\right)  }$.

Whereas the consistency property requires some work to be generalized to TUX
games, the generalizations of the other axioms used by Sobolev are straight-forward.

\medskip

\noindent$2$\emph{-Standardness, }$\mathbf{2}$\textbf{S}$^{\text{\textbf{X}}}%
$\textbf{.}$\;$For all $w\in\mathbb{W}\left(  \left\{  i,j\right\}  \right)
,~\left\{  i,j\right\}  \subseteq\mathbf{U},$ $i\neq j,$ we have%
\[
\varphi_{i}\left(  w\right)  =v\left(  \left\{  i\right\}  ,\left\{  \left\{
j\right\}  \right\}  \right)  +\frac{v\left(  \left\{  i,j\right\}
,\emptyset\right)  -v\left(  \left\{  j\right\}  ,\left\{  \left\{  i\right\}
\right\}  \right)  -v\left(  \left\{  i\right\}  ,\left\{  \left\{  j\right\}
\right\}  \right)  }{2}.
\]

Analogous to
\cite{sobolev1975}%
, we find that the MPW solution is characterized by these properties.

\begin{theorem}
\label{thm:sob+e+2s=sh-star}The MPW solution, $\mathrm{MPW}$, is the unique
solution that satisfies efficiency (\textbf{EF}$\,^{\text{\textbf{X}}}$),
2-standardness ($\mathbf{2}$\textbf{S}$\,^{\text{\textbf{X}}}$), and Sobolev
consistency (\textbf{SC}$^{\text{\textbf{X}}}$).
\end{theorem}

The proof of uniqueness proceeds by induction on the size of the player set
(see \ref{apx:sob+e+2s=sh-star}).

\subsection{Hart and Mas-Colell consistency for TUX games\label{sec:HMconsTUX}%
}

Next, we consider the notion of HM consistency for games with externalities
based on the restriction given in (\ref{eq:r-star}). We define the reduced
game as follows. Let $\varphi$ a be solution for TUX games and let
$w\in\mathbb{W}\left(  N\right)  $, $N\subseteq\mathbf{U}$, and $j\in N$. The
reduced TUX\ game $w_{-j}^{HM,\varphi}\in\mathbb{W}\left(  N\setminus\left\{
j\right\}  \right)  $ is defined by%
\begin{equation}
w_{-j}^{HM,\varphi}\left(  S,\pi\right)  =w\left(  S\cup\left\{  j\right\}
,\pi\right)  -\varphi_{j}\left(  w_{-N\setminus\left(  S\cup\left\{
j\right\}  \right)  }\right)  \label{eq:HMXreducedgame}%
\end{equation}
for all $\left(  S,\pi\right)  \in\mathcal{E}\left(  N\setminus\left\{
j\right\}  \right)  $, where $w_{-N\setminus\left(  S\cup\left\{  j\right\}
\right)  }$ is defined in (\ref{eq:r-star}). This mimics the reduced game for
TU games, where a coalition $S$ can utilize player $j$ but has to compensate
this player according to the solution $\varphi$ in the subgame with player set
$S\cup\left\{  j\right\}  $. This reduction suggests the following property.

\medskip

\noindent\emph{HM Consistency}\textbf{, HMC}$^{\text{\textbf{X}}}$%
\textbf{.}$\;$For all $w\in\mathbb{W}\left(  N\right)  $, $N\subseteq
\mathbf{U},$ and $i,j\in N,$ $i\neq j$, we have $\varphi_{i}\left(  w\right)
=\varphi_{i}(w_{-j}^{HM,\varphi})$.

\medskip

To illustrate this property, recall the public good game in Example
\ref{ex:maskin2003}, for which $\mathrm{MPW}_{1}\left(  w_{-2}\right)  =6.5$,
$\mathrm{MPW}_{1}\left(  w_{-3}\right)  =6$, and $\mathrm{MPW}_{1}\left(
w\right)  =7.5$. The reduced game without player $1$ is given by%
\begin{align*}
w_{-1}^{HM,\varphi}\left(  \left\{  2\right\}  ,\left\{  \left\{  3\right\}
\right\}  \right)   &  =\left(  w\left(  \left\{  2,1\right\}  ,\left\{
\left\{  3\right\}  \right\}  \right)  -\mathrm{MPW}_{1}\left(  w_{-3}\right)
\right)  =6;\\
w_{-1}^{HM,\varphi}\left(  \left\{  3\right\}  ,\left\{  \left\{  2\right\}
\right\}  \right)   &  =\left(  w\left(  \left\{  3,1\right\}  ,\left\{
\left\{  2\right\}  \right\}  \right)  -\mathrm{MPW}_{1}\left(  w_{-2}\right)
\right)  =6.5;\\
w_{-1}^{HM,\varphi}\left(  \left\{  23\right\}  ,\left\{  \emptyset\right\}
\right)   &  =w\left(  \left\{  1,2,3\right\}  ,\emptyset\right)
-\mathrm{MPW}_{1}\left(  w\right)  =16.5.
\end{align*}
Indeed, solving this two-player game gives $\mathrm{MPW}_{2}(w_{-1}%
^{HM,\varphi})=8$ and $\mathrm{MPW}_{3}(w_{-1}^{So,\varphi})=8.5$, which
coincides with the payoffs in the original game $\mathrm{MPW}_{2}\left(
w\right)  $ and $\mathrm{MPW}_{3}\left(  w\right)  $, respectively.

It turns out that the MPW solution can be characterized in a manner similar to
how the Shapley value is characterized using HM consistency.

\begin{theorem}
\label{thm:HMC-star+2s=sh-star}The MPW solution, $\mathrm{MPW}$, is the unique
solution that satisfies 2-standardness ($\mathbf{2}$\textbf{S}%
$\,^{\text{\textbf{X}}}$) and HM consistency (\textbf{HMC}$^{\text{\textbf{X}%
}}$).
\end{theorem}

The proof relies on a different notion of HM consistency, which allows the
removal of multiple players. More precisely, consider the following reduced
TUX\ game $w_{-T}^{HM,\varphi}\in\mathbb{W}\left(  N\setminus T\right)  $
defined by%
\begin{equation}
w_{-T}^{HMC,\varphi}\left(  S,\pi\right)  =w\left(  S\cup T,\pi\right)
-\sum_{j\in T}\varphi_{j}\left(  w_{-N\setminus\left(  S\cup T\right)
}\right)  \label{eq:setHMCXreducedgame}%
\end{equation}
for all $\left(  S,\pi\right)  \in\mathcal{E}\left(  N\setminus T\right)  $,
where $w_{-N\setminus\left(  S\cup T\right)  }\in\mathbb{W}\left(  S\cup
T\right)  $ is defined in (\ref{eq:r-star}). The set version of HM consistency
then reads as follows.

\medskip

\noindent\emph{ Set HM Consistency}\textbf{, SHMC}$^{\text{\textbf{X}}}%
$\textbf{.}$\;$For all $w\in\mathbb{W}\left(  N\right)  $, $N\subseteq
\mathbf{U},$ $i\in N,$ and $T\subseteq N\setminus\left\{  i\right\}  $, we
have $\varphi_{i}\left(  w\right)  =\varphi_{i}(w_{-T}^{HMC,\varphi})$.

\medskip

Note that this boils down to HM consistency with $T=\left\{  j\right\}  $,
i.e., it is obviously not a weaker requirement when allowing the removal of
multiple players. Recall that for TU~games, both notions of HM
consistency---removing one or multiple players---are equivalent
\citep[Lemma~4.4]{HarMas1989}%
. Leveraging the properties of the restriction given in (\ref{eq:r-star}), we
can establish the same for TUX~games.

\begin{proposition}
\label{pro:HMC-star<=>SHMC-star}A solution for TUX\ games $\varphi$ satisfies
HM Consistency (\textbf{HMC}$^{\text{\textbf{X}}}$) if and only if $\varphi$
satisfies Set HM Consistency (\textbf{SHMC}$^{\text{\textbf{X}}}$).
\end{proposition}

We finally mention that
\citet[Theorem~5]{DuEhKa2010}%
\ provide characterizations for alternative solutions for TUX games. These
rely on notions of HM consistency that allow for the removal of multiple
players at once (similar to \textbf{SHMC}$^{\text{\textbf{X}}}$). It remains
an open question whether an equivalence resembling
Proposition~\ref{pro:HMC-star<=>SHMC-star} can be established for their
properties and solutions.

\section{Concluding Remarks}

The widespread application of the Shapley value and its derivatives rest on
its convincing characterizations, in particular, characterizations based on
fairness properties such as balanced contributions by
\citet{myerson1980}%
\ or the consistency properties due to
\cite{sobolev1975}
and
\cite{HarMas1989}%
. Often, solution concepts derived from the Shapley value are motivated by
their derived analogon of the above properties in the specific setup. For
generalizations of the Shapley value to games with externalities (TUX games),
the literature is mainly focused on characterizations that derive from
Shapley's original characterization based on additivity. While technically
attractive, this property is less plausible from a normative perspective.

In this paper, we continued the work of
\citet{DuEhKa2010}
and
\citet{CaFuHu-idd}%
, who investigate TUX games with changing player sets. We demonstrate that the
balanced contributions property can be generalized to TUX games and yields a
characterization of the MPW solution put forth by
\citet{MSPCWe2007}%
. Moreover, we introduce generalizations of Sobolev's consistency and of Hart
and Mas-Colell's consistency to games with externalities (TUX\ games). Again,
this leads to consistency properties that are characteristic of the MPW\ solution.

Characterizations using consistency properties do not only allow us to
distinguish competing solution concepts, but understanding the consistency
property of the MPW solution further aides its computation and implementations
via a mechanism. Moreover, it provides a template of consistency properties of
allocation rules which derive from the MPW solution in specific applications.
Similarly, the balanced contributions property in the presence of
externalities is not only characteristic of the MPW solution, but it likely
remains a plausible requirement for allocation rules that derive from the MPW
solution in specific applications. In this regard, our work may pave the way
for future research on allocation schemes in the presence of externalities for
specific applications such as
\citet{SNCM2023}%
.

\section{Acknowledgements}

Andr\'{e} Casajus: Funded by the Deutsche Forschungsgemeinschaft (DFG, German
Research Foundation) -- 388390901. Yukihiko Funaki: Funded by JSPS KAKENHI
(18KK004, 22H00829).

\begin{small}%
\end{small}%
\appendix

\section{Appendix}

First, we introduce further notation and insights before providing the proofs.
For a given TUX game $w\in\mathbb{W}\left(  N\right)  ,$ $N\subseteq
\mathbf{U}$, we can compute an \textbf{auxiliary TU\ game} $v_{w}^{\star}%
\in\mathbb{V}\left(  N\right)  $, in which each coalition $S$ generates the
worth generated by the grand coalition if we remove all other players
$N\setminus S$ from the game
\citep{DuEhKa2010}%
. That is, the auxiliary TU\ game is defined by%
\begin{equation}
v_{w}^{\star}\left(  S\right)  =w_{-N\setminus S}\left(  S,\emptyset\right)
\qquad\text{for all }S\subseteq N, \label{eq:w-tilde-r}%
\end{equation}
where the restriction $w_{-N\setminus S}$ is given in (\ref{eq:r-star}).

For $N\subseteq\mathbf{U}$ and $\left(  T,\tau\right)  \in\mathcal{E}\left(
N\right)  ,$ the \textbf{scaled Dirac game with externalities}, $\delta
_{T,\tau}\in\mathbb{W}\left(  N\right)  ,$ is defined by%
\begin{equation}
\delta_{T,\tau}\left(  S,\pi\right)  =\left\{
\begin{array}
[c]{ll}%
\frac{1}{p_{N\setminus T}^{\star}\left(  \tau\right)  }, & \text{if }\left(
S,\pi\right)  =\left(  T,\tau\right)  ,\\
0, & \text{else,}%
\end{array}
\right.  \label{eq:scaled-dirac}%
\end{equation}
for all $\left(  S,\pi\right)  \in\mathcal{E}\left(  N\right)  $. Clearly,
these games generalize the Dirac TU\ games. For $N\subseteq\mathbf{U}$ and
$T\subseteq N,$ the \textbf{Dirac TU game} $\delta_{T}^{N}$ is defined by%
\begin{equation}
\delta_{T}^{N}\left(  S\right)  =\left\{
\begin{array}
[c]{ll}%
1, & \text{if }S=T,\\
0, & \text{else}%
\end{array}
\right.  \label{eq:TU-dirac}%
\end{equation}
for all $S\subseteq N.$ Note that every TUX game $w\in\mathbb{W}\left(
N\right)  $ has a unique representation in terms of scaled Dirac games,
\begin{equation}
w=\sum_{\left(  T,\tau\right)  \in\mathcal{E}\left(  N\right)  }w\left(
T,\tau\right)  p_{N\setminus T}^{\star}\left(  \tau\right)  \delta_{T,\tau},
\label{eq:scaled-dirac-rep}%
\end{equation}
i.e., the set $\left\{  \left(  N,\delta_{T,\tau}\right)  \mid\left(
T,\tau\right)  \in\mathcal{E}\left(  N\right)  ,\ T\neq\emptyset\right\}  $ is
a basis of the vector space of all TUX games with player set~$N$.

\subsection{Restriction of scaled Dirac games with
externalities\label{apx:restriction-scaled-dirac}}

For later convenience, we establish the following lemma (note that
\citet[Equation~B.3]{CaFuHu-idd}%
\ establish a similar result for unscaled Dirac games).

\begin{lemma}
\label{lem:restriction-scaled-dirac}The restriction operator $r^{\star}$ is
path independent and for all $N\subseteq\mathbf{U},$ $S\subseteq N,$ and
$\left(  T,\tau\right)  \in\mathcal{E}\left(  N\right)  ,$ we have%
\begin{equation}
\left(  \delta_{T,\tau}\right)  _{-S}=\left\{
\begin{array}
[c]{ll}%
\delta_{T,\tau_{-S}}, & \text{if }S\cap T=\emptyset,\\
\mathbf{0}^{N\setminus S}, & \text{otherwise.}%
\end{array}
\right.  \label{eq:restriction-scaled-dirac}%
\end{equation}

\end{lemma}

\emph{Proof of Lemma \ref{lem:restriction-scaled-dirac}.}\textit{\emph{\/}}
Applying the restriction (\ref{eq:r-star}) gives
\[
\left(  \delta_{T,\tau}\right)  _{-i}\left(  S,\pi\right)  \overset
{\text{(\ref{eq:r-star})}}{=}\frac{1}{n-s}\delta_{T,\tau}\left(
S,\pi_{+i\rightsquigarrow\emptyset}\right)  +\frac{\left\vert B\right\vert
}{n-s}\sum_{B\in\pi}\delta_{T,\tau}\left(  S,\pi_{+i\rightsquigarrow
B}\right)  .
\]
By definition of $\delta_{T,\tau}$, (\ref{eq:restriction-scaled-dirac}), the
right-hand side of the upper equation vanishes unless $\left(  S,\pi
_{+i\rightsquigarrow\emptyset}\right)  =\left(  T,\tau\right)  $ or $\left(
S,\pi_{+i\rightsquigarrow B}\right)  =\left(  T,\tau\right)  $, for when it
becomes $\delta_{T,\tau}\left(  T,\tau\right)  /\left(  n-t\right)  $ or
$\left\vert B\right\vert \delta_{T,\tau}\left(  T,\tau\right)  /\left(
n-t\right)  $, respectively. We get%
\[
\left(  \delta_{T,\tau}\right)  _{-i}\left(  S,\pi\right)  \overset
{\text{(\ref{eq:restriction-scaled-dirac})}}{=}\left\{
\begin{array}
[c]{ll}%
\frac{1}{n-t}\frac{1}{p_{N\setminus T}^{\star}\left(  \tau\right)  }, &
\text{if }i\in N\setminus T\text{, }\left(  S,\pi\right)  =\left(  T,\tau
_{-i}\right)  \text{, and }\tau\left(  i\right)  =\left\{  i\right\}  ,\\
\frac{\left\vert \tau\left(  i\right)  \right\vert }{n-t}\frac{1}%
{p_{N\setminus T}^{\star}\left(  \tau\right)  } & \text{if }i\in N\setminus
T\text{, }\left(  S,\pi\right)  =\left(  T,\tau_{-i}\right)  \text{, and
}\left\vert \tau\left(  i\right)  \right\vert >1,\\
0, & \text{if }i\in T.
\end{array}
\right.
\]
By definition of $p^{\star}$, (\ref{eq:p-star}),
\[
\frac{1}{p_{\left(  N\setminus\left\{  i\right\}  \right)  \setminus T}%
^{\star}\left(  \tau_{-i}\right)  }=\left\{
\begin{array}
[c]{ll}%
\frac{1}{n-t}\frac{1}{p_{N\setminus T}^{\star}\left(  \tau\right)  }, &
\text{if }i\in N\setminus T\text{ and }\tau\left(  i\right)  =\left\{
i\right\}  ,\\
\frac{\left\vert \tau\left(  i\right)  \right\vert }{n-t}\frac{1}%
{p_{N\setminus T}^{\star}\left(  \tau\right)  }, & \text{if }i\in N\setminus
T\text{ and }\left\vert \tau\left(  i\right)  \right\vert >1.
\end{array}
\right.
\]
Therefore, if $i\in N\setminus T$, then $\left(  \delta_{T,\tau}\right)
_{-i}\left(  S,\pi\right)  =\delta_{T,\tau_{-i}}\left(  S,\pi\right)  $; and
if $i\in T$, then $\left(  \delta_{T,\tau}\right)  _{-i}\left(  S,\pi\right)
=0$. This means we obtain%
\begin{equation}
\delta_{T,\tau}=\left\{
\begin{array}
[c]{ll}%
\delta_{T,\tau_{-i}}, & \text{if }i\in N\setminus T,\\
\mathbf{0}^{N\setminus\left\{  i\right\}  }, & \text{if }i\in T.
\end{array}
\right.  \label{eq:restriction-scaled-Dirac-1}%
\end{equation}

Notice that $\left(  \left(  \delta_{T,\tau}\right)  _{-i}\right)
_{-j}=\mathbf{0}^{N\setminus\left\{  i,j\right\}  }$ if $i\in T$ or $j\in T$;
and that $\left(  \left(  \delta_{T,\tau}\right)  _{-i}\right)  _{-j}%
=\delta_{T,\tau_{-\left\{  i,j\right\}  }}=\left(  \left(  \delta_{T,\tau
}\right)  _{-j}\right)  _{-i}$ if $i,j\in N\setminus T$. Hence, restriction
operator $r^{\star}$ is path independent for scaled Dirac games. Since these
constitute a basis of the space of TUX\ games and since the restriction
operator $r^{\star}$ is linear, this holds true for all TUX games (see
\citet[Theorem~7]{CaFuHu-idd}%
\ for an alternative proof using unscaled Dirac games). In particular, the
order in which players are removed from $\delta_{T,\tau}$ does not matter, the
removal of a coalition $S\subseteq N$ is well-defined, and the claim follows
from repeated application of (\ref{eq:restriction-scaled-Dirac-1}).

\subsection{Average and auxiliary TU\ game of scaled Dirac games with
externalities\label{apx:avg-dirac-aux}}

We next establish the following Lemma (note that
\citet[Equation~B.5]{CaFuHu-idd}%
\ establish a similar result for unscaled Dirac games).

\begin{lemma}
\label{lem:avg-dirac-aux}For all $N\subseteq\mathbf{U}$ and $\left(
T,\tau\right)  \in\mathcal{E}\left(  N\right)  ,$ we have%
\begin{equation}
\bar{v}_{\delta_{T,\tau}}=\delta_{T}^{N}=v_{\delta_{T,\tau}}^{\star}%
\in\mathbb{V}\left(  N\right)  . \label{eq:avg-dirac-aux}%
\end{equation}

\end{lemma}

\emph{Proof of Lemma \ref{lem:avg-dirac-aux}.}\textit{\emph{\/}} For
$N\subseteq\mathbf{U}$, $\left(  T,\tau\right)  \in\mathcal{E}\left(
N\right)  ,$ and $S\subseteq N,$ we immediately get from the definition of the
average game $\bar{v}$, (\ref{eq:averagegame}), and the definition of the
scaled Dirac games with externalities $\delta_{T,\tau}$,
(\ref{eq:scaled-dirac}), that the factors $p_{N\setminus T}^{\star}\left(
\tau\right)  $ cancel out, and we have
\[
\bar{v}_{\delta_{T,\tau}}\overset{\text{(\ref{eq:averagegame}%
),(\ref{eq:p-star}),(\ref{eq:scaled-dirac})}}{=}\delta_{T}^{N}.
\]
For $N\subseteq\mathbf{U}$ and $\left(  T,\tau\right)  \in\mathcal{E}\left(
N\right)  ,$ using the definition of the auxiliary game $v^{\star}$,
(\ref{eq:w-tilde-r}), the definition of the scaled Dirac games with
externalities $\delta_{T,\tau}$, (\ref{eq:scaled-dirac}),
Lemma~\ref{lem:restriction-scaled-dirac}, and $p_{\emptyset}^{\star}\left(
\emptyset\right)  =1$ gives%
\begin{align*}
v_{\delta_{T,\tau}}^{\star}\left(  S\right)   &  \overset
{\text{(\ref{eq:w-tilde-r})}}{=}\left(  \delta_{T,\tau}\right)  _{-N\setminus
S}\left(  S,\emptyset\right) \\
&  \overset{\text{Lemma (\ref{lem:restriction-scaled-dirac})}%
,\text{(\ref{eq:scaled-dirac})}}{=}\left\{
\begin{array}
[c]{ll}%
1, & \text{if }T=S\text{ (and }\tau=\emptyset\text{),}\\
\mathbf{0}^{N\setminus S}, & \text{otherwise.}%
\end{array}
\right. \\
&  =\delta_{T}^{N}\left(  S\right)  .
\end{align*}

\begin{remark}
The lemma implies $v_{w}^{\star}=\bar{v}_{w}$ for all $w\in\mathbb{W}\left(
N\right)  ,$ $N\subseteq\mathbf{U}.$ This follows from linearity of both the
average operator and the restriction operator as well as the uniqueness of the
coefficients in (\ref{eq:scaled-dirac-rep}). Further, we have%
\[
\mathrm{MPW}\left(  w\right)  \overset{\text{(\ref{eq:Sh-star})}}%
{=}\mathrm{Sh}\left(  \bar{v}_{w}\right)  =\mathrm{Sh}\left(  v_{w}^{\star
}\right)  \qquad\text{for all }w\in\mathbb{W}\left(  N\right)  ,~N\subseteq
\mathbf{U}.
\]

\end{remark}

\subsection{Proof of Lemma~\ref{lem:r-and-avg-commute}}

The lemma follows from linearity of the involved operators, the fact that
scaled Dirac games with externalities constitute a basis of $\mathbb{W}$, and
from the following: if $i\notin T$ then%
\[
\left(  \bar{v}_{\delta_{T,\tau}}\right)  _{-i}\overset
{\text{(\ref{eq:avg-dirac-aux})}}{=}\left(  \delta_{T}^{N}\right)
_{-i}=\delta_{T}^{N\setminus\left\{  i\right\}  }\overset
{\text{(\ref{eq:avg-dirac-aux})}}{=}\bar{v}_{\delta_{T,\tau_{-i}}}%
\overset{\text{(\ref{eq:restriction-scaled-dirac})}}{=}\bar{v}_{\left(
\delta_{T,\tau}\right)  _{-i}},
\]
whereas everything being the null game if $i\in T$.

\subsection{Proof of Theorem~\ref{thm:BCx+EFx<=>sh-star}}

\emph{Existence:}\textit{\emph{\/}} We need to show that the MPW solution
satisfies \textbf{BC}$^{\text{\textbf{X}}}$. By the definition of the MPW
solution (\ref{eq:Sh-star}), i.e., $\mathrm{MPW}_{i}\left(  w\right)
=\mathrm{Sh}_{i}\left(  \bar{v}_{w}\right)  ,$ and commutation of the average
operator and of and removal operator as shown in
Lemma~\ref{lem:r-and-avg-commute}, i.e., $\bar{v}_{w_{-j}}=\left(  \bar{v}%
_{w}\right)  _{-j}$, we get%
\begin{equation}
\mathrm{MPW}_{i}\left(  w_{-j}\right)  \overset{\text{(\ref{eq:Sh-star})}}%
{=}\mathrm{Sh}_{i}\left(  \bar{v}_{w_{-j}}\right)  \overset
{\text{Lemma~\ref{lem:r-and-avg-commute}}}{=}\mathrm{Sh}_{i}\left(  \left(
\bar{v}_{w}\right)  _{-j}\right)  . \label{eq:proof-thm-bcx}%
\end{equation}
Now we can confirm that the MPW solution inherits \textbf{BC}%
$^{\text{\textbf{X}}}$ from the Shapley value,
\begin{multline*}
\mathrm{MPW}_{i}\left(  w\right)  -\mathrm{MPW}_{i}\left(  w_{-j}\right)
\overset{\text{(\ref{eq:proof-thm-bcx}),(\ref{eq:Sh-star})}}{=}\mathrm{Sh}%
_{i}\left(  \bar{v}_{w}\right)  -\mathrm{Sh}_{i}\left(  \left(  \bar{v}%
_{w}\right)  _{-j}\right) \\
\overset{\text{\textbf{BC} of }\mathrm{Sh}\text{ }}{=}\mathrm{Sh}_{j}\left(
\bar{v}_{w}\right)  -\mathrm{Sh}_{j}\left(  \left(  \bar{v}_{w}\right)
_{-i}\right)  \overset{\text{(\ref{eq:proof-thm-bcx}),(\ref{eq:Sh-star})}}%
{=}\mathrm{MPW}_{j}\left(  w\right)  -\mathrm{MPW}_{j}\left(  w_{-i}\right)  .
\end{multline*}

\emph{Uniqueness:}\textit{\emph{\/}} Let $\varphi$ satisfy and \textbf{EF}%
$^{\text{\textbf{X}}}$ and \textbf{BC}$^{\text{\textbf{X}}}$. We show
$\varphi=\mathrm{MPW}$ by induction on $n$. \emph{Induction basis}%
\textit{\emph{\/}}: For $n=1,$ the claim $\varphi=\mathrm{MPW}$ is immediate
from \textbf{EF}$^{\text{\textbf{X}}}$. \emph{Induction hypothesis\textit{\/}
}\textbf{(}\emph{IH}\textit{\emph{\/}}\textbf{)}: $\varphi\left(  w\right)
=\mathrm{MPW}\left(  w\right)  $ for all $w\in\mathbb{W}\left(  N\right)  $
such that $n\leq\ell$. \emph{Induction step}\textit{\emph{\/}}: Let
$w\in\mathbb{W}\left(  N\right)  $ such that $n=\ell+1.$ By \textbf{BC}%
$^{\text{\textbf{X}}},$ we have%
\[
\varphi_{i}\left(  w\right)  -\varphi_{j}\left(  w\right)  =\varphi_{i}\left(
w_{-j}\right)  -\varphi_{j}\left(  w_{-i}\right)  \qquad\text{for all }i,j\in
N.
\]
Summing up the equation over all $j\in N$ gives%
\[
n\varphi_{i}\left(  w\right)  -\sum_{j\in N}\varphi_{j}\left(  w\right)
=\sum_{j\in N}\left(  \varphi_{i}\left(  w_{-j}\right)  -\varphi_{j}\left(
w_{-i}\right)  \right)  \overset{\emph{IH}}{=}\sum_{j\in N}\left(
\mathrm{MPW}_{i}\left(  w_{-j}\right)  -\mathrm{MPW}_{j}\left(  w_{-i}\right)
\right)  .
\]
Finally, applying \textbf{EF}$^{\text{\textbf{X}}}$ gives
\[
n\varphi_{i}\left(  w\right)  =\sum_{j\in N}\left(  \mathrm{MPW}_{i}\left(
w_{-j}\right)  -\mathrm{MPW}_{j}\left(  w_{-i}\right)  \right)  +w\left(
N,\emptyset\right)  ,
\]
that is, $\varphi_{i}\left(  w\right)  $ is uniquely determined for all $i\in
N$. This implies $\varphi_{i}\left(  w\right)  =\mathrm{MPW}_{i}\left(
w\right)  .$

\subsection{\label{apx:sob+e+2s=sh-star}Proof of
Theorem~\ref{thm:sob+e+2s=sh-star}}

\emph{Existence\textit{\/}}: We need to show that the MPW solution satisfies
\textbf{SC}$^{\text{\textbf{X}}}$. To this end, we compute the average game of
the reduced game $w_{-j}^{So,\mathrm{MPW}}$,
\[
\bar{v}_{w_{-j}^{So,\mathrm{MPW}}}\left(  S\right)  \overset
{\text{(\ref{eq:sobolev-TUX-reduced-game}),(\ref{eq:averagegame})}}{=}%
\sum_{\pi\in\Pi\left(  N\setminus S\right)  }p^{\star}\left(  \pi\right)
\left(  \frac{s}{n-1}\left(  w\left(  S\cup\left\{  j\right\}  ,\pi\right)
-\mathrm{MPW}_{j}\left(  w\right)  \right)  +\frac{n-s-1}{n-1}w_{-j}\left(
S,\pi\right)  \right)  .
\]
For $\left(  S,\pi\right)  \in\mathcal{E}\left(  N\setminus\left\{  j\right\}
\right)  ,$ we obtain%
\begin{align*}
\sum_{\pi\in\Pi\left(  N\setminus S\right)  }p^{\star}\left(  \pi\right)
w\left(  S\cup\left\{  j\right\}  ,\pi\right)   &  \overset
{\text{(\ref{eq:averagegame})}}{=}\bar{v}_{w}\left(  S\cup\left\{  j\right\}
\right)  ,\\
\sum_{\pi\in\Pi\left(  N\setminus S\right)  }p^{\star}\left(  \pi\right)
\mathrm{MPW}_{j}\left(  w\right)   &  =\mathrm{Sh}_{j}\left(  \bar{v}%
_{w}\right)  ,\text{ and}\\
\sum_{\pi\in\Pi\left(  N\setminus S\right)  }p^{\star}\left(  \pi\right)
w_{-j}\left(  S,\pi\right)   &  \overset{\text{(\ref{eq:averagegame})}}{=}%
\bar{v}_{w_{-j}}\left(  S\right)  \overset{\text{Lemma
\ref{lem:r-and-avg-commute}}}{=}\left(  \bar{v}_{w}\right)  _{-j}\left(
S\right)  =\bar{v}_{w}\left(  S\right)  ,
\end{align*}
where we used the definition of the average game $\bar{v}_{w}$
(\ref{eq:averagegame}); the fact that $p^{\star}$ is a probability
distribution and the definition $\mathrm{MPW}_{j}\left(  w\right)
=\mathrm{Sh}_{j}\left(  \bar{v}_{w}\right)  $ (\ref{eq:Sh-star}); and finally
the commutation Lemma \ref{lem:r-and-avg-commute}. We get
\begin{equation}
\bar{v}_{w_{-j}^{So,\mathrm{MPW}}}\left(  S\right)  =\frac{s}{n-1}\left(
\bar{v}_{w}\left(  S\cup\left\{  j\right\}  \right)  -\mathrm{Sh}_{j}\left(
\bar{v}_{w}\right)  \right)  +\frac{n-s-1}{n-1}\bar{v}_{w}\left(  S\right)
=\left(  \bar{v}_{w}\right)  _{-j}^{So,\mathrm{Sh}}\left(  S\right)  .
\label{eq:proofSobCon}%
\end{equation}
Applying the Shapley value on both sides and using Sobolev consistency
(\textbf{SC}) of the Shapley value gives%
\[
\mathrm{Sh}_{i}\left(  \bar{v}_{w_{-j}^{So,\mathrm{MPW}}}\right)
\overset{\text{(\ref{eq:proofSobCon})}}{=}\mathrm{Sh}_{i}\left(  \left(
\bar{v}_{w}\right)  _{-j}^{So,\mathrm{Sh}}\right)  \overset{\text{\textbf{SC}
of }\mathrm{Sh}}{=}\mathrm{Sh}_{i}\left(  \bar{v}_{w}\right)
\]
for all $i\in N\setminus j$. By definition of $\mathrm{MPW}$ (\ref{eq:Sh-star}%
), the LHS is just $\mathrm{MPW}_{i}(w_{-j}^{So,\mathrm{MPW}})$ and the
RHS\ equals $\mathrm{MPW}_{i}\left(  w\right)  $, which establishes the claim.

\emph{Uniqueness\textit{\/}}: Let the $\varphi$ satisfy \textbf{EF}%
$^{\text{\textbf{X}}}$, \textbf{2ST}$^{\text{\textbf{X}}}$, and \textbf{SC}%
$^{\text{\textbf{X}}}$. We show $\varphi=\mathrm{MPW}$ by induction on $n.$
\emph{Induction basis\/}: For $n\leq2,$the claim is immediate from
\textbf{EF}$^{\text{\textbf{X}}}$ and \textbf{2ST}$^{\text{\textbf{X}}}$.
\emph{Induction hypothesis\textit{\/} }\textbf{(}\emph{IH}\textit{\emph{\/}%
}\textbf{)}: $\varphi\left(  w\right)  =\mathrm{MPW}\left(  w\right)  $ for
all $w\in\mathbb{W}\left(  N\right)  ,$ $N\subseteq\mathbf{U}$ such that
$n\leq\ell,$ $\ell\geq2.$ \emph{Induction step}\/: Let $w\in\mathbb{W}\left(
N\right)  $ be such that $n=\ell+1.$ For $i\in N,$ fix $k\in N\setminus
\left\{  i\right\}  .$ We obtain%
\[
\varphi_{i}\left(  w\right)  \overset{\text{\textbf{SC}}^{\text{\textbf{X}}}%
}{=}\varphi_{i}\left(  w_{-k}^{So,\varphi}\right)  \overset{\emph{IH}}%
{=}\mathrm{MPW}_{i}\left(  w_{-k}^{So,\varphi}\right)  \overset
{\text{\textbf{SC}}^{\text{\textbf{X}}}}{=}\mathrm{MPW}_{i}\left(  w\right)
.
\]

\subsection{HM-reduction and player removal can be
swapped\label{apx:HM-red-res-commutes}}

We prepare the proof of Proposition~\ref{pro:HMC-star<=>SHMC-star} by showing
that it does not matter whether we first compute the HM-reduced game $\left(
w_{-T}^{HM,\varphi}\right)  _{-S}$ or whether we first remove players,
$\left(  w_{-S}\right)  _{-T}^{HM,\varphi}$.

\begin{lemma}
\label{lem:HM-red-res-commutes}For all TUX games $w\in\mathbb{W}\left(
N\right)  $, $N\subseteq\mathbf{U}$, and coalitions $S,T\in N$, such that
$S\cap T=\emptyset,$ we have%
\[
\left(  w_{-T}^{HM,\varphi}\right)  _{-S}=\left(  w_{-S}\right)
_{-T}^{HM,\varphi}.
\]

\end{lemma}

\emph{Proof of Lemma \ref{lem:HM-red-res-commutes}.}\textit{\emph{\/}} We
first show that%
\begin{equation}
w_{-S}\left(  R,\rho\right)  =\sum_{\tau\in\Pi\left(  N\setminus R\right)
:\tau_{-S}=\rho}\frac{p_{N\setminus T}^{\star}\left(  \tau\right)
}{p_{\left(  N\setminus S\right)  \setminus R}^{\star}\left(  \rho\right)
}w\left(  R,\tau\right)  \label{eq:star-gen}%
\end{equation}
for all $w\in\mathbb{W}\left(  N\right)  $, $N\subseteq\mathbf{U}$,
$S\subseteq N$ and $\left(  R,\rho\right)  \in\mathcal{E}\left(  N\setminus
S\right)  .$ With the linearity of the restriction given in (\ref{eq:r-star}),
we can exploit the representation of games by scaled Dirac games
(\ref{eq:scaled-dirac-rep}), that is, we have%
\[
w_{-S}=\sum_{\left(  T,\tau\right)  \in\mathcal{E}\left(  N\right)  }w\left(
T,\tau\right)  p_{N\setminus T}^{\star}\left(  \tau\right)  \left(
\delta_{T,\tau}\right)  _{-S}%
\]
for all $w\in\mathbb{V}$ and $S\subseteq N$. Applying the formula for
restrictions of scaled Dirac games (\ref{eq:restriction-scaled-dirac}) then
gives%
\[
w_{-S}=\sum_{\left(  T,\tau\right)  \in\mathcal{E}\left(  N\right)  :S\cap
T=\emptyset}w\left(  T,\tau\right)  p_{N\setminus T}^{\star}\left(
\tau\right)  \delta_{T,\tau_{-S}}.
\]
The evaluation of $w_{-S}$ at $\left(  R,\rho\right)  \in\mathcal{E}\left(
N\setminus S\right)  $ vanishes if $\left(  R,\rho\right)  \neq\left(
T,\tau_{-S}\right)  $. For $\left(  R,\rho\right)  =\left(  T,\tau
_{-S}\right)  $, it becomes $1/p_{\left(  N\setminus S\right)  \setminus
R}^{\star}\left(  \rho\right)  $, so that%
\[
w_{-S}\left(  R,\rho\right)  =\sum_{\tau\in\Pi\left(  N\setminus R\right)
:\tau_{-S}=\rho}p_{N\setminus T}^{\star}\left(  \tau\right)  w\left(
R,\tau\right)  \frac{1}{p_{\left(  N\setminus S\right)  \setminus R}^{\star
}\left(  \rho\right)  },
\]
which reduces to the formula in the lemma by the definition of $p^{\star}$
(\ref{eq:p-star}).

Let now $R\subseteq N\setminus\left(  S\cup T\right)  $ and $\rho\in\Pi\left(
\left(  N\setminus\left(  S\cup T\right)  \right)  \setminus R\right)  $. We
obtain%
\begin{align*}
&  \left(  w_{-T}^{HM,\varphi}\right)  _{-S}\left(  R,\rho\right) \\
&  \overset{\text{(\ref{eq:star-gen})}}{=}\sum_{\tau\in\Pi\left(  \left(
N\setminus T\right)  \setminus R\right)  :\tau_{-S}=\rho}\frac{p_{\left(
N\setminus T\right)  \setminus R}^{\star}\left(  \tau\right)  }{p_{\left(
\left(  N\setminus T\right)  \setminus S\right)  \setminus R}^{\star}\left(
\rho\right)  }w_{-T}^{HM,\varphi}\left(  R,\tau\right) \\
&  \overset{\text{(\ref{eq:setHMCXreducedgame})}}{=}\sum_{\tau\in\Pi\left(
\left(  N\setminus T\right)  \setminus R\right)  :\tau_{-S}=\rho}%
\frac{p_{\left(  N\setminus T\right)  \setminus R}^{\star}\left(  \tau\right)
}{p_{\left(  \left(  N\setminus T\right)  \setminus S\right)  \setminus
R}^{\star}\left(  \rho\right)  }\left(  w\left(  R\cup T,\tau\right)
-\sum_{j\in T}\varphi_{j}\left(  w_{-N\setminus\left(  R\cup T\right)
}\right)  \right)  .
\end{align*}
On the other hand, we have%
\begin{align*}
&  \left(  w_{-S}\right)  _{-T}^{HM,\varphi}\left(  R,\rho\right) \\
&  \overset{\text{(\ref{eq:setHMCXreducedgame})}}{=}w_{-S}\left(  R\cup
T,\rho\right)  -\sum_{j\in T}\varphi_{j}\left(  w_{-N\setminus\left(  R\cup
T\right)  }\right) \\
&  \overset{\text{(\ref{eq:star-gen})}}{=}\sum_{\tau\in\Pi\left(
N\setminus\left(  R\cup T\right)  \right)  :\tau_{-S}=\rho}\frac
{p_{N\setminus\left(  R\cup T\right)  }^{\star}\left(  \tau\right)
}{p_{\left(  N\setminus S\right)  \setminus\left(  R\cup T\right)  }^{\star
}\left(  \rho\right)  }\left(  w\left(  R\cup T,\tau\right)  -\sum_{j\in
T}\varphi_{j}\left(  w_{-N\setminus\left(  R\cup T\right)  }\right)  \right)
.
\end{align*}
Since $R\subseteq N\setminus\left(  S\cup T\right)  $, and since $S$ and $T$
are disjoint, both expressions are the same.

\subsection{Proof of Proposition~\ref{pro:HMC-star<=>SHMC-star}}

It is clear that \textbf{SHMC}$^{\text{\textbf{X}}}$ implies \textbf{HMC}%
$^{\text{\textbf{X}}}$. Let the solution $\varphi$ satisfy \textbf{HMC}%
$^{\text{\textbf{X}}}$. We show that $\varphi$ satisfies \textbf{SHMC}%
$^{\text{\textbf{X}}}$ by induction on the size of the removed coalition $T.$

\emph{Induction basis}\textit{\emph{\/}}: For $t=1,$ the claim is immediate by
\textbf{HMC}$^{\text{\textbf{X}}}$.

\emph{Induction hypothesis\textit{\/} }\textbf{(}\emph{IH}\textit{\emph{\/}%
}\textbf{)}: \textbf{SHMC}$^{\text{\textbf{X}}}$ implies \textbf{HMC}%
$^{\text{\textbf{X}}}$ for $t=\ell$.

\emph{Induction step}\textit{\emph{\/}}: Let $w\in\mathbb{W}\left(  N\right)
,$ and let $T\subseteq N$ such that $t=\ell+1.$ For $i\in N\setminus T$ and
$j\in T,$ we get%
\[
\varphi_{i}\left(  v\right)  \overset{\emph{IH}}{=}\varphi_{i}\left(
w_{-T\setminus\left\{  j\right\}  }^{HM,\varphi}\right)  \overset
{\text{\textbf{HMC}}^{\text{\textbf{X}}}}{=}\varphi_{i}\left(  \left(
w_{-T\setminus\left\{  j\right\}  }^{HM,\varphi}\right)  _{-j}^{HM,\varphi
}\right)  .
\]
Now, it suffices to show that $(w_{-T\setminus\left\{  j\right\}
}^{HM,\varphi})_{-j}^{HM,\varphi}\left(  S,\pi\right)  =w_{-T}^{HM,\varphi
}\left(  S,\pi\right)  $ for $S\subseteq N\setminus T$ and $\pi\in\Pi\left(
\left(  N\setminus T\right)  \setminus S\right)  $ in order to establish
\textbf{SHMC}$^{\text{\textbf{X}}}$. Indeed, using the notation $\tilde
{S}=\left(  N\setminus\left(  T\setminus\left\{  j\right\}  \right)  \right)
\setminus\left(  S\cup\left\{  j\right\}  \right)  =N\setminus\left(  S\cup
T\right)  ,$ we get the following two equations:%
\begin{align*}
(w_{-T\setminus\left\{  j\right\}  }^{HM,\varphi})_{-j}^{HM,\varphi}\left(
S,\pi\right)   &  \overset{\text{(\ref{eq:setHMCXreducedgame})}}%
{=}w_{-T\setminus\left\{  j\right\}  }^{HM,\varphi}\left(  S\cup\left\{
j\right\}  ,\pi\right)  -\varphi_{j}((w_{-T\setminus\left\{  j\right\}
}^{HM,\varphi})_{-\tilde{S}})\\
&  \overset{\text{(\ref{eq:setHMCXreducedgame})}}{=}w\left(  S\cup
T,\pi\right)  -\sum_{k\in T\setminus\left\{  j\right\}  }\varphi
_{k}(w_{-\tilde{S}})-\varphi_{j}((w_{-T\setminus\left\{  j\right\}
}^{HM,\varphi})_{-\tilde{S}});
\end{align*}
and%
\begin{align*}
w_{-T}^{HM,\varphi}\left(  S,\pi\right)   &  \overset
{\text{(\ref{eq:setHMCXreducedgame})}}{=}w\left(  S\cup T,\pi\right)
-\sum_{k\in T\setminus\left\{  j\right\}  }\varphi_{k}\left(  w_{-\tilde{S}%
}\right)  -\varphi_{j}\left(  w_{-\tilde{S}}\right) \\
&  \overset{\emph{IH}}{=}w\left(  S\cup T,\pi\right)  -\sum_{k\in
T\setminus\left\{  j\right\}  }\varphi_{k}(w_{-\tilde{S}})-\varphi
_{j}((w_{-\tilde{S}})_{-T\setminus\left\{  j\right\}  }^{HM,\varphi}).
\end{align*}
Thus, we need to show that the order of removal and reduction does not matter,
that is, we need to show
\[
\left(  w_{-T\setminus\left\{  j\right\}  }^{HM,\varphi}\right)  _{-\tilde{S}%
}=\left(  w_{-\tilde{S}}\right)  _{-T\setminus\left\{  j\right\}
}^{HM,\varphi}.
\]
Indeed, this is true because of Lemma~\ref{lem:HM-red-res-commutes}. Note that
we used the specifics of the restriction given in (\ref{eq:r-star}) only in
the last step.

\subsection{Proof of Theorem~\ref{thm:HMC-star+2s=sh-star}}

\emph{Existence:}\textit{\emph{\/}} It is well-known that $\mathrm{MPW}$
inherits \textbf{2ST}$^{\text{\textbf{X}}}$ from $\mathrm{Sh}$. To establish
\textbf{HMC}$\,^{\text{\textbf{X}}}$ of $\mathrm{MPW}$, consider the average
game of the reduced game $w_{-j}^{HM,\mathrm{MPW}}$,%
\[
\bar{v}_{w_{-j}^{HM,\mathrm{MPW}}}\left(  S\right)  \overset
{\text{(\ref{eq:HMXreducedgame}),(\ref{eq:averagegame})}}{=}=\sum_{\pi\in
\Pi\left(  N\setminus S\right)  }p^{\star}\left(  \pi\right)  \left(  w\left(
S\cup\left\{  j\right\}  ,\pi\right)  -\mathrm{MPW}_{j}\left(  w_{-N\setminus
\left(  S\cup\left\{  j\right\}  \right)  }\right)  \right)  .
\]
for $\left(  S,\pi\right)  \in\mathcal{E}\left(  N\setminus\left\{  j\right\}
\right)  .$ We can insert%
\begin{align*}
\sum_{\pi\in\Pi\left(  N\setminus S\right)  }p^{\star}\left(  \pi\right)
w\left(  S\cup\left\{  j\right\}  ,\pi\right)   &  =\bar{v}_{w}\left(
S\cup\left\{  j\right\}  \right)  ,\text{ and}\\
\sum_{\pi\in\Pi\left(  N\setminus S\right)  }p^{\star}\left(  \pi\right)
\mathrm{MPW}_{j}\left(  w_{-N\setminus\left(  S\cup\left\{  j\right\}
\right)  }\right)   &  =\mathrm{Sh}_{j}\left(  \bar{v}_{w_{-N\setminus\left(
S\cup\left\{  j\right\}  \right)  }}\right)  ,
\end{align*}
giving
\begin{equation}
\bar{v}_{w_{-j}^{HM,\mathrm{MPW}}}\left(  S\right)  =\bar{v}_{w}\left(
S\cup\left\{  j\right\}  \right)  -\mathrm{Sh}_{j}\left(  \bar{v}%
_{w_{-N\setminus\left(  S\cup\left\{  j\right\}  \right)  }}\right)  =\left(
\bar{v}_{w}\right)  _{-j}^{HM,\mathrm{Sh}}\left(  S\right)
\label{eq:proofHMCon}%
\end{equation}
for all $S\subseteq N\setminus\left\{  j\right\}  $. Applying the Shapley
value on both sides and using HM consistency (\textbf{HMC}) of the Shapley
value gives%
\[
\mathrm{Sh}_{i}\left(  \bar{v}_{w_{-j}^{HM,\mathrm{MPW}}}\right)
\overset{\text{(\ref{eq:proofHMCon})}}{=}\mathrm{Sh}_{i}\left(  \left(
\bar{v}_{w}\right)  _{-j}^{HM,\mathrm{Sh}}\right)  \overset{\text{\textbf{HMC}
of }\mathrm{Sh}}{=}\mathrm{Sh}_{i}\left(  N,\bar{v}_{w}\right)  .
\]
for all $i\in N\setminus j$. By definition of $\mathrm{MPW}$ (\ref{eq:Sh-star}%
), the LHS is just $\mathrm{MPW}_{i}(w_{-j}^{HM,\mathrm{MPW}})$ and the
RHS\ equals $\mathrm{MPW}_{i}\left(  w\right)  $, which establishes the claim.

\emph{Uniqueness:}\textit{\emph{\/}} Let $\varphi$ satisfy \textbf{HMC}%
$\,^{\text{\textbf{X}}}$ and \textbf{2ST}$^{\text{\textbf{X}}}$. By
Proposition~\ref{pro:HMC-star<=>SHMC-star}, $\varphi$ satisfies \textbf{SHMC}%
$\,^{\text{\textbf{X}}}.$

We first establish that $\varphi$ is efficient (\textbf{EF}$^{\text{\textbf{X}%
}}$). To this end, we proceed by induction on the size of the player set $n$.
\emph{Induction basis}\textit{\emph{\/}}: For $n=2,$ the claim is immediate
from \textbf{2ST}$^{\text{\textbf{X}}}$, and for $n=1$ it further follows from
\textbf{SHMC}$^{\text{\textbf{X}}}$. \emph{Induction hypothesis\textit{\/}
}\textbf{(}\emph{IH}\textit{\emph{\/}}\textbf{)}: \textbf{SHMC}%
$\,^{\text{\textbf{X}}}$ and \textbf{2ST}$^{\text{\textbf{X}}}$ together imply
\textbf{EF}$^{\text{\textbf{X}}}$ for $n<\ell$. \emph{Induction step}%
\textit{\emph{\/}}: Let $w\in\mathbb{W}\left(  N\right)  $ with $n=\ell.$
Then, by definition of the reduced game (\ref{eq:setHMCXreducedgame}) for
$T=N\setminus\left\{  i\right\}  $, we have%
\[
w_{-N\setminus\left\{  i\right\}  }^{HM,\varphi}\left(  \left\{  i\right\}
,\pi\right)  =w\left(  N,\pi\right)  -\sum_{k\in N\setminus\left\{  i\right\}
}\varphi_{k}\left(  w\right)  .
\]
On the other hand, we have%
\[
w_{-N\setminus\left\{  i\right\}  }^{HM,\varphi}\left(  \left\{  i\right\}
,\pi\right)  \overset{\emph{IH}}{=}\varphi_{i}\left(  w_{-N\setminus\left\{
i\right\}  }^{HMC,\varphi}\right)  \overset{\text{\textbf{SHMC}}%
^{\text{\textbf{X}}}}{=}\varphi_{i}\left(  w\right)  .
\]
The two equations together confirm \textbf{EF}$^{\text{\textbf{X}}}.$

To establish uniqueness, i.e., $\varphi=\mathrm{MPW}$, we again proceed by
induction on $n$. \emph{Induction basis}\textit{\emph{\/}}: For $n=2,$ the
claim is immediate from \textbf{2ST}$^{\text{\textbf{X}}}$, and for $n=1$ it
further follows from \textbf{EF}$^{\text{\textbf{X}}}$. \emph{Induction
hypothesis\textit{\/} }\textbf{(}\emph{IH}\textit{\emph{\/}}\textbf{)}:
$\varphi=\mathrm{MPW}$ for $n<\ell,$ where $\ell>2$. \emph{Induction
step}\textit{\emph{\/}}: Let $w\in\mathbb{W}\left(  N\right)  $ with
$n=\ell>2.$ Fix two players $i,k\in N$ and consider the reduced games on these
two players $w_{-N\setminus\left\{  i,k\right\}  }^{HM,\varphi}$ and
$w_{-N\setminus\left\{  i,k\right\}  }^{HM,\mathrm{MPW}}$. Note that%
\begin{align*}
w_{-N\setminus\left\{  i,k\right\}  }^{HM,\varphi}\left(  \left\{  i\right\}
,\pi\right)   &  =w\left(  N\setminus\left\{  k\right\}  ,\pi\right)
-\sum_{j\in N\setminus\left\{  i,k\right\}  }\varphi_{j}\left(  w_{-k}\right)
\\
&  \overset{\text{\emph{IH}}}{=}w\left(  N\setminus\left\{  k\right\}
,\pi\right)  -\sum_{j\in N\setminus\left\{  i,k\right\}  }\mathrm{MPW}%
_{j}\left(  w_{-k}\right)  =w_{-N\setminus\left\{  i,k\right\}  }%
^{HM,\mathrm{MPW}}\left(  \left\{  i\right\}  ,\pi\right)  .
\end{align*}
Analogously, $w_{-N\setminus\left\{  i,k\right\}  }^{HM,\varphi}\left(
\left\{  k\right\}  ,\pi\right)  =w_{-N\setminus\left\{  i,k\right\}
}^{HM,\mathrm{MPW}}\left(  \left\{  k\right\}  ,\pi\right)  $. By
\textbf{2ST}$^{\text{\textbf{X}}}$, the differences of payoffs of $i$ and $k$
must equal, i.e.,%
\begin{align*}
&  \varphi_{i}\left(  w_{-N\setminus\left\{  i,k\right\}  }^{HM,\varphi
}\right)  -\varphi_{k}\left(  w_{-N\setminus\left\{  i,k\right\}
}^{HM,\varphi}\right) \\
&  \overset{\text{\textbf{2ST}}^{\text{\textbf{X}}}\text{ of }\varphi}%
{=}w_{-N\setminus\left\{  i,k\right\}  }^{HM,\varphi}\left(  \left\{
i\right\}  ,\pi\right)  -w_{-N\setminus\left\{  i,k\right\}  }^{HM,\varphi
}\left(  \left\{  k\right\}  ,\pi\right) \\
&  =w_{-N\setminus\left\{  i,k\right\}  }^{HM,\mathrm{MPW}}\left(  \left\{
i\right\}  ,\pi\right)  -w_{-N\setminus\left\{  i,k\right\}  }%
^{HM,\mathrm{MPW}}\left(  \left\{  k\right\}  ,\pi\right) \\
&  \overset{\text{\textbf{2ST}}^{\text{\textbf{X}}}\text{ of }\mathrm{MPW}}%
{=}\mathrm{MPW}_{i}\left(  w_{-N\setminus\left\{  i,k\right\}  }^{HM,\varphi
}\right)  -\mathrm{MPW}_{k}\left(  w_{-N\setminus\left\{  i,k\right\}
}^{HM,\varphi}\right)  .
\end{align*}
Applying \textbf{SHMC}$^{\text{\textbf{X}}}$ further augments this to the
original game,%
\[
\varphi_{i}\left(  w\right)  -\varphi_{k}\left(  w\right)  =\mathrm{MPW}%
_{i}\left(  w\right)  -\mathrm{MPW}_{k}\left(  w\right)
\]
for arbitrary $i,k\in N.$ Summing up this equation over all $k\in N$ and
applying \textbf{EF}$^{\text{\textbf{X}}}$ finally gives $\varphi_{i}\left(
w\right)  =\mathrm{MPW}_{i}\left(  w\right)  $ for all $i\in N$.

\bibliographystyle{elsart-harv}
\bibliography{mpwsob}

@Preamble{"\newcommand{\noopsort}[1]{}"}

@Article{AlArRu2005,
  author    = {M. J. Albizuri and J. Arin and J. Rubio},
  title     = {An axiom system for a value for games in partition function form},
  journal   = {International Game Theory Review},
  year      = {2005},
  volume    = {7},
  pages     = {63-73},
}

@Article{CliSer2008,
  author    = {Geoffroy {\noopsort{Clippel}}{de Clippel} and Roberto Serrano},
  title     = {Marginal contributions and externalities in the value},
  journal   = {Econometrica},
  year      = {2008},
  volume    = {76},
  number    = {6},
  pages     = {1413-1436},
}

@Article{DuEhKa2010,
  author  = {Bhaskar Dutta and Lars Ehlers and Anirban Kar},
  title   = {Externalities, potential, value and consistency},
  journal = {Journal of Economic Theory},
  year    = {2010},
  volume  = {145},
  number  = {6},
  pages   = {2380-2411},
  issn    = {0022-0531},
}

@Article{bolger1989,
  author  = {E. M. Bolger},
  title   = {A Set of Axioms for a Value for Partition Function Games},
  journal = {International Journal of Game Theory},
  year    = {1989},
  volume  = {18},
  pages   = {37-44},
}

@Article{ThrLuc1963,
  author  = {R. M. Thrall and W. F. Lucas},
  title   = {$n$-Person Games in Partition Function Form},
  journal = {Naval Research Logistic Quarterly},
  year    = {1963},
  volume  = {10},
  pages   = {281-293},
}

@Article{MSPCWe2007,
  author       = {Macho-Stadler, I. and P{\'e}rez-Castrillo, D. and Wettstein, D.},
  title        = {Sharing the surplus: An extension of the {Shapley} value for environments with externalities},
  journal      = {Journal of Economic Theory},
  journaltitle = {Journal of Economic Theory},
  year         = {2007},
  volume       = {135},
  number       = {1},
  pages        = {339-356},
}

@Article{McQSug2018,
  author       = {Ben McQuillin and Robert Sugden},
  title        = {Balanced externalities and the {Shapley} value},
  journal      = {Games and Economic Behavior},
  journaltitle = {Games and Economic Behavior},
  year         = {2018},
  volume       = {108},
  pages        = {81-92},
}

@Article{mcquillin2009,
  author    = {Ben McQuillin},
  title     = {The extended and generalized {Shapley} value: {Simultaneous} consideration of coalitional externalities and coalitional structure},
  journal   = {Journal of Economic Theory},
  year      = {2009},
  volume    = {144},
  pages     = {696-721},
}

@Article{young1985,
  author  = {H. P. Young},
  title   = {Monotonic Solutions of Cooperative Games},
  journal = {International Journal of Game Theory},
  year    = {1985},
  volume  = {14},
  pages   = {65-72},
}

@Article{SkMiWo2018,
  author       = {Oskar Skibski and Thomasz P. Michalak and Michael Woolbridge},
  title        = {The stochastic {Shapley} value for coalitional games with externalities},
  journal      = {Games and Economic Bahavior},
  journaltitle = {Games and Economic Bahavior},
  year         = {2018},
  volume       = {108},
  pages        = {65-80},
}

@InCollection{shapley1953,
  author    = {Lloyd S. Shapley},
  title     = {A value for $n$-person games},
  booktitle = {Contributions to the Theory of Games},
  year      = {1953},
  editor    = {H.W. Kuhn and A.W. Tucker},
  volume    = {II},
  publisher = {Princeton University Press},
  pages     = {307-317},
  address   = {Princeton},
}

@Article{PhaNor2007,
  author    = {K. H. {Pham Do} and H. Norde},
  title     = {The {Shapley} value for partition function form games},
  journal   = {International Game Theory Review},
  year      = {2007},
  volume    = {9},
  pages     = {353-360},
}

@Article{ewens1972,
  author    = {Warren Ewens},
  title     = {The sampling theory of selectively neutral alleles},
  journal   = {Theoretical Population Biology},
  year      = {1972},
  volume    = {3},
  pages     = {87-112},
}

@InCollection{aldous1985,
  author    = {D. Aldous},
  title     = {Exchangeability and related topics},
  booktitle = {Ecole d'été de probabilités de Saint-Flour, XIII---1983},
  year      = {1985},
  editor    = {P. L. Hennequin},
  volume    = {1117},
  series    = {Lecture Notes in Mathematics},
  publisher = {Springer},
  chapter   = {1},
  pages     = {1-198},
  address   = {Berlin},
}

@Article{HarMas1989,
  author    = {Sergiu Hart and A. {Mas-Colell}},
  title     = {Potential, value, and consistency},
  journal   = {Econometrica},
  year      = {1989},
  volume    = {57},
  number    = {3},
  pages     = {589-614},
}

@Article{myerson1980,
  author  = {Roger B. Myerson},
  title   = {Conference Structures and Fair Allocation Rules},
  journal = {International Journal of Game Theory},
  year    = {1980},
  volume  = {9},
  pages   = {169-182},
}

@Article{CasHue-deva,
  author       = {André Casajus and Frank Huettner},
  title        = {Decomposition of solutions and the {Shapley} value},
  journal      = {Games and Economic Behavior},
  journaltitle = {Games and Economic Behavior},
  year         = {2018},
  volume       = {13},
  number       = {3},
  pages        = {1-23},
}

@InBook{sobolev1975,
  author       = {A. I. Sobolev},
  pages        = {95-151},
  publisher    = {Academy of Sciences of the Lithuanian SSR},
  title        = {The characterization of optimality principles in cooperative games by functional equations},
  year         = {1975},
  address      = {Vilnius},
  volume       = {6},
  bookauthor   = {N. N. Vorobiev},
  booktitle    = {Mathematical Methods in the Social Sciences},
  origlanguage = {Russian},
}

@Article{casajus-potential,
  author    = {André Casajus},
  title     = {Potential, value, and random partitions},
  journal   = {Economics Letters},
  year      = {2014},
  volume    = {126},
  number    = {2},
  pages     = {164-166},
}

@Article{SkiMic2020,
  author  = {Oskar Skibski and Tomasz Michalak},
  title   = {Fair division in the presence of externalities},
  doi     = {10.1007/s00182-019-00682-4},
  number  = {1},
  pages   = {147--172},
  volume  = {49},
  journal = {International Journal of Game Theory},
  month   = {may},
  year    = {2019},
}

@Article{crane2016,
  author    = {Harry Crane},
  journal   = {Statistical Science},
  title     = {The Ubiquitous Ewens Sampling Formula},
  year      = {2016},
  number    = {1},
  pages     = {1--19},
  volume    = {31},
  doi       = {10.1214/15-sts529},
  publisher = {Institute of Mathematical Statistics},
}

@Book{PelSud2007,
  author    = {Bezalel Peleg and Peter Sudhölter},
  publisher = {Springer-Verlag GmbH},
  title     = {Introduction to the Theory of Cooperative Games},
  year      = {2007},
  isbn      = {9783540729457},
  month     = aug,
  ean       = {9783540729457},
}

@Article{McQSug2016,
  author  = {Ben McQuillin and Robert Sugden},
  title   = {Backward Induction Foundations of the Shapley Value},
  doi     = {10.3982/ecta13191},
  number  = {6},
  pages   = {2265--2280},
  volume  = {84},
  journal = {Econometrica},
  year    = {2016},
}

@Article{CaFuHu-idd,
  author  = {Casajus, André and Funaki, Yukihiko and Huettner, Frank},
  journal = {Games and Economic Behavior},
  title   = {Random partitions, potential, value, and externalities},
  year    = {2024},
  pages   = {88-106},
  volume  = {147},
  doi     = {doi.org/10.1016/j.geb.2024.06.004},
}

@Article{driessen1991,
  author    = {Theo S. H. Driessen},
  title     = {A Survey of Consistency Properties in Cooperative Game Theory},
  doi       = {10.1137/1033003},
  number    = {1},
  pages     = {43-59},
  volume    = {33},
  journal   = {{SIAM} Review},
  publisher = {Society for Industrial {\&} Applied Mathematics ({SIAM})},
  year      = {1991},
}

@Article{aumann_maschler1985,
  author  = {Aumann, R. and Maschler, M.},
  journal = {Journal of Economic Theory},
  title   = {{G}ame {T}heoretic {A}nalysis of a {B}ankruptcy {P}roblem from the {T}almud},
  year    = {1985},
  pages   = {195-213},
  volume  = {36},
}

@Article{oneill1982,
  author    = {Barry O'Neill},
  journal   = {Mathematical Social Sciences},
  title     = {A problem of rights arbitration from the Talmud},
  year      = {1982},
  number    = {4},
  pages     = {345-371},
  volume    = {2},
  doi       = {10.1016/0165-4896(82)90029-4},
  publisher = {Elsevier {BV}},
}

@Article{PotSud1999,
  author    = {Jos Potters and Peter Sudhölter},
  title     = {Airport problems and consistent allocation rules},
  doi       = {10.1016/s0165-4896(99)00004-9},
  number    = {1},
  pages     = {83-102},
  volume    = {38},
  journal   = {Mathematical Social Sciences},
  publisher = {Elsevier {BV}},
  year      = {1999},
}

@Article{LitOwe1973,
  author    = {S. C. Littlechild and G. Owen},
  journal   = {Management Science},
  title     = {A Simple Expression for the Shapely Value in a Special Case},
  year      = {1973},
  issn      = {00251909, 15265501},
  number    = {3},
  pages     = {370--372},
  volume    = {20},
  publisher = {INFORMS},
  urldate   = {2022-08-04},
}

@Article{SudZar2017,
  author    = {Peter Sudhölter and Jos{\'{e}} M. Zarzuelo},
  journal   = {European Journal of Operational Research},
  title     = {Characterizations of highway toll pricing methods},
  year      = {2017},
  number    = {1},
  pages     = {161-170},
  volume    = {260},
  doi       = {10.1016/j.ejor.2016.11.051},
  publisher = {Elsevier {BV}},
}

@Article{KuMoZa2013,
  author    = {Jeroen Kuipers and Manuel A. Mosquera and Jos{\'{e}} M. Zarzuelo},
  journal   = {European Journal of Operational Research},
  title     = {Sharing costs in highways: A game theoretic approach},
  year      = {2013},
  number    = {1},
  pages     = {158-168},
  volume    = {228},
  doi       = {10.1016/j.ejor.2013.01.018},
  publisher = {Elsevier {BV}},
}

@Article{XuDrSuSu2013,
  author    = {Genjiu Xu and Theo S.H. Driessen and Hao Sun and Jun Su},
  title     = {Consistency for the additive efficient normalization of semivalues},
  doi       = {10.1016/j.ejor.2012.08.018},
  number    = {3},
  pages     = {566-571},
  volume    = {224},
  journal   = {European Journal of Operational Research},
  publisher = {Elsevier {BV}},
  year      = {2013},
}

@Article{GraFun2012,
  author    = {Michel Grabisch and Yukihiko Funaki},
  title     = {A coalition formation value for games in partition function form},
  doi       = {10.1016/j.ejor.2012.02.036},
  number    = {1},
  pages     = {175-185},
  volume    = {221},
  journal   = {European Journal of Operational Research},
  publisher = {Elsevier {BV}},
  year      = {2012},
}

@Article{myerson1977pffg,
  author  = {Roger B. Myerson},
  journal = {International Journal of Game Theory},
  title   = {Values for Games in Partition Function Form},
  year    = {1977},
  pages   = {23-31},
  volume  = {6},
}

@Article{BeGrSa2023,
  author    = {Gustavo Berganti{\~{n}}os and Carlos Groba and Antonio Sartal},
  title     = {Applying the Shapley value to the tuna fishery},
  doi       = {10.1016/j.ejor.2022.12.040},
  number    = {1},
  pages     = {306-318},
  volume    = {309},
  journal   = {European Journal of Operational Research},
  publisher = {Elsevier {BV}},
  year      = {2023},
}

@Article{BeVP2010,
  author    = {Gustavo Berganti{\~{n}}os and Juan Vidal-Puga},
  journal   = {European Journal of Operational Research},
  title     = {Realizing fair outcomes in minimum cost spanning tree problems through non-cooperative mechanisms},
  year      = {2010},
  number    = {3},
  pages     = {811-820},
  volume    = {201},
  doi       = {10.1016/j.ejor.2009.04.003},
  publisher = {Elsevier {BV}},
}

@Article{JuChBr2014,
  author    = {Yuan Ju and Youngsub Chun and Ren{\'{e}} van den Brink},
  journal   = {Journal of Economic Theory},
  title     = {Auctioning and selling positions: A non-cooperative approach to queueing conflicts},
  year      = {2014},
  pages     = {33-45},
  volume    = {153},
  doi       = {10.1016/j.jet.2014.05.007},
  publisher = {Elsevier {BV}},
}

@Article{winter1992,
  author    = {Eyal Winter},
  journal   = {Games and Economic Behavior},
  title     = {The consistency and potential for values of games with coalition structure},
  year      = {1992},
  number    = {1},
  pages     = {132-144},
  volume    = {4},
  doi       = {10.1016/0899-8256(92)90009-h},
  publisher = {Elsevier {BV}},
}

@Article{GoGrGr2021,
  author    = {Sanjith Gopalakrishnan and Daniel Granot and Frieda Granot},
  title     = {Consistent Allocation of Emission Responsibility in Fossil Fuel Supply Chains},
  doi       = {10.1287/mnsc.2020.3874},
  number    = {12},
  pages     = {7637-7668},
  volume    = {67},
  journal   = {Management Science},
  publisher = {Institute for Operations Research and the Management Sciences ({INFORMS})},
  year      = {2021},
}

@Article{maniquet2003,
  author    = {François Maniquet},
  journal   = {Journal of Economic Theory},
  title     = {A characterization of the {Shapley} value in queueing problems},
  year      = {2003},
  pages     = {90-103},
  volume    = {109},
}

@Article{BenHav2018,
  author    = {Dan Bendel and Moshe Haviv},
  title     = {Cooperation and sharing costs in a tandem queueing network},
  doi       = {10.1016/j.ejor.2018.04.049},
  number    = {3},
  pages     = {926-933},
  volume    = {271},
  journal   = {European Journal of Operational Research},
  publisher = {Elsevier {BV}},
  year      = {2018},
}

@Article{CuPeTi1989,
  author    = {Imma Curiel and Giorgio Pederzoli and Stef Tijs},
  title     = {Sequencing games},
  doi       = {10.1016/0377-2217(89)90427-x},
  number    = {3},
  pages     = {344-351},
  volume    = {40},
  journal   = {European Journal of Operational Research},
  publisher = {Elsevier {BV}},
  year      = {1989},
}

@Article{DavMas1965,
  author    = {Morton Davis and Michael Maschler},
  title     = {The kernel of a cooperative game},
  doi       = {10.1002/nav.3800120303},
  number    = {3},
  pages     = {223-259},
  volume    = {12},
  journal   = {Naval Research Logistics Quarterly},
  publisher = {Wiley},
  year      = {1965},
}

@ARTICLE{thomson2015,
  author = {William Thomson},
  title = {Axiomatic and game-theoretic analysis of bankruptcy and taxation
	problems: {An} update},
  journal = {Mathematical Social Sciences},
  year = {2015},
  volume = {74},
  pages = {41-59}
}

@Article{BriChu2011,
  author    = {Ren{\'{e}} van den Brink and Youngsub Chun},
  journal   = {Social Choice and Welfare},
  title     = {Balanced consistency and balanced cost reduction for sequencing problems},
  year      = {2011},
  month     = {apr},
  number    = {3},
  pages     = {519--529},
  volume    = {38},
  doi       = {10.1007/s00355-011-0533-6},
  publisher = {Springer Science and Business Media {LLC}},
}

@Article{ThoVel2022,
  author    = {William Thomson and Rodrigo A. Velez},
  journal   = {Mathematical Programming},
  title     = {Consistent queueing rules},
  year      = {2022},
  month     = {dec},
  doi       = {10.1007/s10107-022-01905-5},
  publisher = {Springer Science and Business Media {LLC}},
}

@Article{DutKar2004,
  author    = {Bhaskar Dutta and Anirban Kar},
  journal   = {Games and Economic Behavior},
  title     = {Cost monotonicity, consistency and minimum cost spanning tree games},
  year      = {2004},
  number    = {2},
  pages     = {223--248},
  volume    = {48},
  doi       = {10.1016/j.geb.2003.09.008},
  publisher = {Elsevier {BV}},
}

@Article{chambers2004,
  author    = {Christopher P. Chambers},
  journal   = {Journal of Mathematical Economics},
  title     = {Consistency in the probabilistic assignment model},
  year      = {2004},
  number    = {8},
  pages     = {953--962},
  volume    = {40},
  doi       = {10.1016/j.jmateco.2003.10.004},
  publisher = {Elsevier {BV}},
}

@Article{LipCon2001,
  author    = {Lipovetsky, Stan and Conklin, Michael},
  journal   = {Applied Stochastic Models in Business and Industry},
  title     = {Analysis of Regression in Game Theory Approach},
  year      = {2001},
  number    = {4},
  pages     = {319--330},
  volume    = {17},
  outdoi    = {10.1002/asmb.446},
  publisher = {Wiley},
}

@article{Shorrocks2012,
  title = {Decomposition Procedures for Distributional Analysis: A Unified Framework Based on the {{Shapley}} Value},
  author = {Shorrocks, Anthony F.},
  year = {2012},
  journal = {The Journal of Economic Inequality},
  volume = {11},
  number = {1},
  pages = {99--126},
  publisher = {{Springer Science and Business Media LLC}},
  outdoi = {10.1007/s10888-011-9214-z}
}

@Article{LunLee2017,
  author  = {Lundberg, Scott M and Lee, Su-In},
  journal = {Advances in neural information processing systems},
  title   = {A Unified Approach to Interpreting Model Predictions},
  year    = {2017},
  volume  = {30},
}

@Article{Lundbergetal2020,
  author    = {Lundberg, Scott M. and Erion, Gabriel and Chen, Hugh and DeGrave, Alex and Prutkin, Jordan M. and Nair, Bala and Katz, Ronit and Himmelfarb, Jonathan and Bansal, Nisha and Lee, Su-In},
  journal   = {Nature Machine Intelligence},
  title     = {From Local Explanations to Global Understanding with Explainable {{AI}} for Trees},
  year      = {2020},
  number    = {1},
  pages     = {56--67},
  volume    = {2},
  outdoi    = {10.1038/s42256-019-0138-9},
  publisher = {{Springer Science and Business Media LLC}},
}

@Article{Gul1989,
  author    = {Gul, Faruk},
  journal   = {Econometrica : journal of the Econometric Society},
  title     = {Bargaining Foundations of Shapely Value},
  year      = {1989},
  number    = {1},
  pages     = {81},
  volume    = {57},
  outdoi    = {10.2307/1912573},
  publisher = {JSTOR},
}

@Article{BruGauMen2018,
  author    = {Br{\"u}gemann, Bj{\"o}rn and Gautier, Pieter and Menzio, Guido},
  journal   = {The Review of Economic Studies},
  title     = {Intra Firm Bargaining and {{Shapley}} Values},
  year      = {2018},
  number    = {2},
  pages     = {564--592},
  volume    = {86},
  outdoi    = {10.1093/restud/rdy015},
  publisher = {Oxford University Press (OUP)},
}

@Article{fujinaka2004,
  author  = {Yuji Fujinaka},
  journal = {mimeo},
  title   = {On the Marginality Principle in Partition Function Form Games},
  year    = {2004},
}

@Article{BBRW2021,
  author  = {Basso, Franco and Basso, Leonardo J. and Rönnqvist, Mikael and Weintraub, Andres},
  journal = {European Journal of Operational Research},
  title   = {Coalition formation in collaborative production and transportation with competing firms},
  year    = {2021},
  issn    = {0377-2217},
  number  = {2},
  pages   = {569-581},
  volume  = {289},
  doi     = {10.1016/j.ejor.2020.07.039},
}

@Article{SNCM2023,
  author  = {Saavedra-Nieves, Alejandro and Casas-Méndez, Balbina},
  journal = {European Journal of Operational Research},
  title   = {On the centrality analysis of covert networks using games with externalities},
  year    = {2023},
  issn    = {0377-2217},
  number  = {3},
  pages   = {1365-1378},
  volume  = {309},
  doi     = {10.1016/j.ejor.2023.02.023},
}

@Article{AMoEhl-prenuk,
  author  = {{\noopsort{Alvarez}}Álvarez-Mozos, Mikel and Ehlers, Lars},
  journal = {Mathematical Social Sciences},
  title   = {Externalities and the (pre)nucleolus in cooperative games},
  year    = {2024},
  pages   = {10-15},
  volume  = {128},
  doi     = {10.1016/j.mathsocsci.2024.01.003},
}

@Article{LFAMCMHe2007,
  author  = {Lorenzo-Freire, Silvia and Alonso-Meijide, José M. and Casas-Méndez, Balbina and Hendrickx, Ruud},
  journal = {European Journal of Operational Research},
  title   = {Balanced contributions for TU games with awards and applications},
  year    = {2007},
  number  = {2},
  pages   = {958-964},
  volume  = {182},
  doi     = {10.1016/j.ejor.2006.07.040},
}

@Article{GRVP2010,
  author  = {Gómez-Rúa, María and Vidal-Puga, Juan},
  journal = {European Journal of Operational Research},
  title   = {The axiomatic approach to three values in games with coalition structure},
  year    = {2010},
  number  = {2},
  pages   = {795-806},
  volume  = {207},
  doi     = {10.1016/j.ejor.2010.05.014},
}

@Article{MGPC2023,
  author  = {Meng, Fan-Yong and Gong, Zai-Wu and Pedrycz, Witold and Chu, Jun-Fei},
  journal = {European Journal of Operational Research},
  title   = {Selfish-dilemma consensus analysis for group decision making in the perspective of cooperative game theory},
  year    = {2023},
  number  = {1},
  pages   = {290-305},
  volume  = {308},
  doi     = {10.1016/j.ejor.2022.12.019},
}

@Article{DPGJCM2022,
  author  = {Davila-Pena, Laura and García-Jurado, Ignacio and Casas-Méndez, Balbina},
  journal = {European Journal of Operational Research},
  title   = {Assessment of the influence of features on a classification problem: An application to COVID-19 patients},
  year    = {2022},
  number  = {2},
  pages   = {631-641},
  volume  = {299},
  doi     = {10.1016/j.ejor.2021.09.027},
}

@Article{GAMaPo2015,
  author  = {González Arangüena, Enrique and Manuel, Conrado Miguel and del Pozo, Mónica},
  journal = {European Journal of Operational Research},
  title   = {Values of games with weighted graphs},
  year    = {2015},
  number  = {1},
  pages   = {248-257},
  volume  = {243},
  doi     = {10.1016/j.ejor.2014.11.033},
}

@Article{BrGAMaPo2014,
  author  = {van den Brink, René and González-Arangüena, Enrique and Manuel, Conrado and del Pozo, Mónica},
  journal = {European Journal of Operational Research},
  title   = {Order monotonic solutions for generalized characteristic functions},
  year    = {2014},
  number  = {3},
  pages   = {786-796},
  volume  = {238},
  doi     = {10.1016/j.ejor.2014.04.016},
}

@Article{BerMT2023,
  author        = {Gustavo Bergantiños and Juan D. Moreno-Ternero},
  journal       = {Management Science},
  title         = {Revenue sharing at music streaming platforms},
  year          = {2025},
  number        = {forthcoming},
  archiveprefix = {arXiv},
  doi           = {https://arxiv.org/abs/2310.11861},
  eprint        = {2310.11861},
}

@Article{KamKon2012,
  author    = {Y. Kamijo and T. Kongo},
  journal   = {European Journal of Operational Research},
  title     = {Whose deletion does not affect your payoff? {The} difference between the {Shapley} value, the egalitarian value, the solidarity value, and the {Banzhaf} value},
  year      = {2012},
  pages     = {638-646},
  volume    = {216},
}

@Book{moulin2003,
  author    = {Moulin, Herv{\'e}},
  publisher = {MIT press},
  title     = {Fair division and collective welfare},
  year      = {2003},
}

@Article{hafalir2007,
  author  = {Hafalir, Isa E.},
  journal = {Games and Economic Behavior},
  title   = {Efficiency in coalition games with externalities},
  year    = {2007},
  number  = {2},
  pages   = {242-258},
  volume  = {61},
  doi     = {10.1016/j.geb.2007.01.009},
}

@Article{yi1997,
  author  = {Yi, Sang-Seung},
  journal = {Games and Economic Behavior},
  title   = {Stable Coalition Structures with Externalities},
  year    = {1997},
  number  = {2},
  pages   = {201-237},
  volume  = {20},
  doi     = {10.1006/game.1997.0567},
}

@InProceedings{maskin2003,
  author       = {Maskin, Eric},
  title        = {Bargaining, Coalitions and Externalities},
  year         = {2003},
  address      = {Princeton},
  note         = {Presidential Address, Institute for Advanced Study},
  organization = {Econometric Society},
}

@PhdThesis{feldman1996,
  author = {Feldman, B. E.},
  school = {State University of New York at Stony Brook},
  title  = {Bargaining, Coalition Formation, and Value},
  year   = {1996},
}

@Book{Pitman2006,
  author    = {Pitman, Jim},
  publisher = {Springer},
  title     = {Combinatorial Stochastic Processes},
  year      = {2006},
  isbn      = {9783540309901},
  pages     = {256},
  subtitle  = {Ecole d'Eté de Probabilités de Saint-Flour XXXII - 2002 (Lecture Notes in Mathematics 1875)},
}

\end{document}